\documentclass{article}
\usepackage{csquotes}  
\usepackage{arxiv}
\usepackage[pdftex]{graphicx}
\usepackage{amsmath}
\usepackage{enumerate}
\usepackage{mcode}   
\usepackage{array} % and/or
\usepackage{longtable} % and/or
\usepackage{colortab} % or
\usepackage{colortbl}
\usepackage{arydshln}
\usepackage{amsthm}
\usepackage{hyperref}
\usepackage{import}
\usepackage{tikz}
\usepackage{bm}
\usepackage[colorinlistoftodos]{todonotes}
\usepackage{algorithm}
\usepackage{algpseudocode}
\usetikzlibrary{matrix,decorations.pathreplacing, calc, positioning}
\newtheorem{theorem}{Theorem}

\newtheorem*{remark}{Remark}

\begin{document}
\title{The Multi-Source Preemptive $M/PH/1/1$ Queue 
	with Packet Errors: Exact Distribution of the Age of Information and Its Peak}
\author{
	Ozancan~Doğan\\
	Electrical and Electronics Engineering Dept.\\
	Bilkent University, 06800\\
	Ankara, Turkey \\
	\texttt{ozancan@ee.bilkent.edu.tr} \\
	\And
    Nail~Akar\\
	Electrical and Electronics Engineering Dept.\\
	Bilkent University, 06800\\
	Ankara, Turkey \\
	\texttt{akar@ee.bilkent.edu.tr} \\
	 }
	\footnote{Mr. Dogan is supported in part by the {\em 5G and Beyond} scholarship granted by the Information and Communication Technologies Authority (ICTA) of Turkey and Vodafone Turkey. }
\maketitle
\begin{abstract}
	Age of Information (AoI) and Peak AoI (PAoI) and their analytical models have recently drawn substantial amount of attention in information theory and wireless communications disciplines, in the context of qualitative assessment of information freshness in status update systems. We take a queueing-theoretic approach and study a probabilistically preemptive bufferless $M/PH/1/1$ queueing system with arrivals stemming from $N$ separate information sources, with the aim of modeling a generic status update system. In this model, a new information packet arrival from source $m$ is allowed to preempt a packet from source $n$ in service, with a probability depending on $n$ and $m$. To make the model even more general than the existing ones, for each of the information sources, we assume a distinct PH-type service time distribution and a distinct packet error probability.
	Subsequently, we obtain the exact distributions of the AoI and PAoI for each of the information sources using matrix-analytical algorithms and in particular the theory of Markov fluid queues and sample path arguments. This is in contrast with existing methods that rely on Stochastic Hybrid Systems (SHS) which obtain only the average values and in less general settings.   
	Numerical examples are provided to validate the proposed approach as well as to give engineering insight on the impact of preemption probabilities on certain AoI and PAoI performance figures. 
\end{abstract}

\section{Introduction}
\label{intro}
Timely status updates are key for stable operation in networked control and monitoring systems. 
Lately, there has been substantial amount of interest centered around Age of Information (AoI) and Peak AoI (PAoI) processes
in the fields of information theory and wireless communications  
in the context of qualitative assessment of information freshness in status update systems \cite{kaul_etal_SMAN11,kaul_etal_infocom12,kaul_etal_ciss12,pappas_etal_icc15,kosta_etal,kosta_etal_survey,sun_etal_tit17,yates_kaul_tit19}.
The survey \cite{kosta_etal_survey} provides a relatively recent overview of the AoI concept and its applications.
AoI performance-related studies include those that propose analytical models for AoI \cite{inoue_etal_tit19,costa_etal_TIT16,chen_huang_isit16} or research focusing on optimization of AoI-related performance metrics \cite{sun_etal_tit17,huang_modiano,arafa_ulukus_asilomar17,hsu_etal_isit17,he_etal_TIT18}. 
Actually, the AoI metric keeps track of the staleness of a remote monitor's knowledge of a stochastic process randomly sampled and transmitted (in the form of information packets) by an information source where the monitor and the source reside at two separate points in a packet-based communications network.  More formally, the AoI maintained at the monitor for a given source is defined as the time elapsed since the generation of the last successfully received update packet. Consequently, the AoI process turns out to be a cyclic process that increases in time with unit slope within a cycle with the exception that the AoI process undergoes abrubt downward jumps at random status packet reception epochs. 
After such a jump, a new cycle begins. Under continuous-time stationary scenarios, the AoI process is a stationary continuous-time, continuous-valued stochastic process for which our interest in this paper is in finding its exact marginal distribution in a specific scenario to be detailed.  
A related equally important process is the discrete-time, continuous-valued PAoI process that is obtained by taking the peak values during each cycle of the AoI process \cite{costa_peak}, the exact distribution of which is also sought in this paper.
\begin{figure}[tb]
	\centering
	\includegraphics[width=\linewidth]{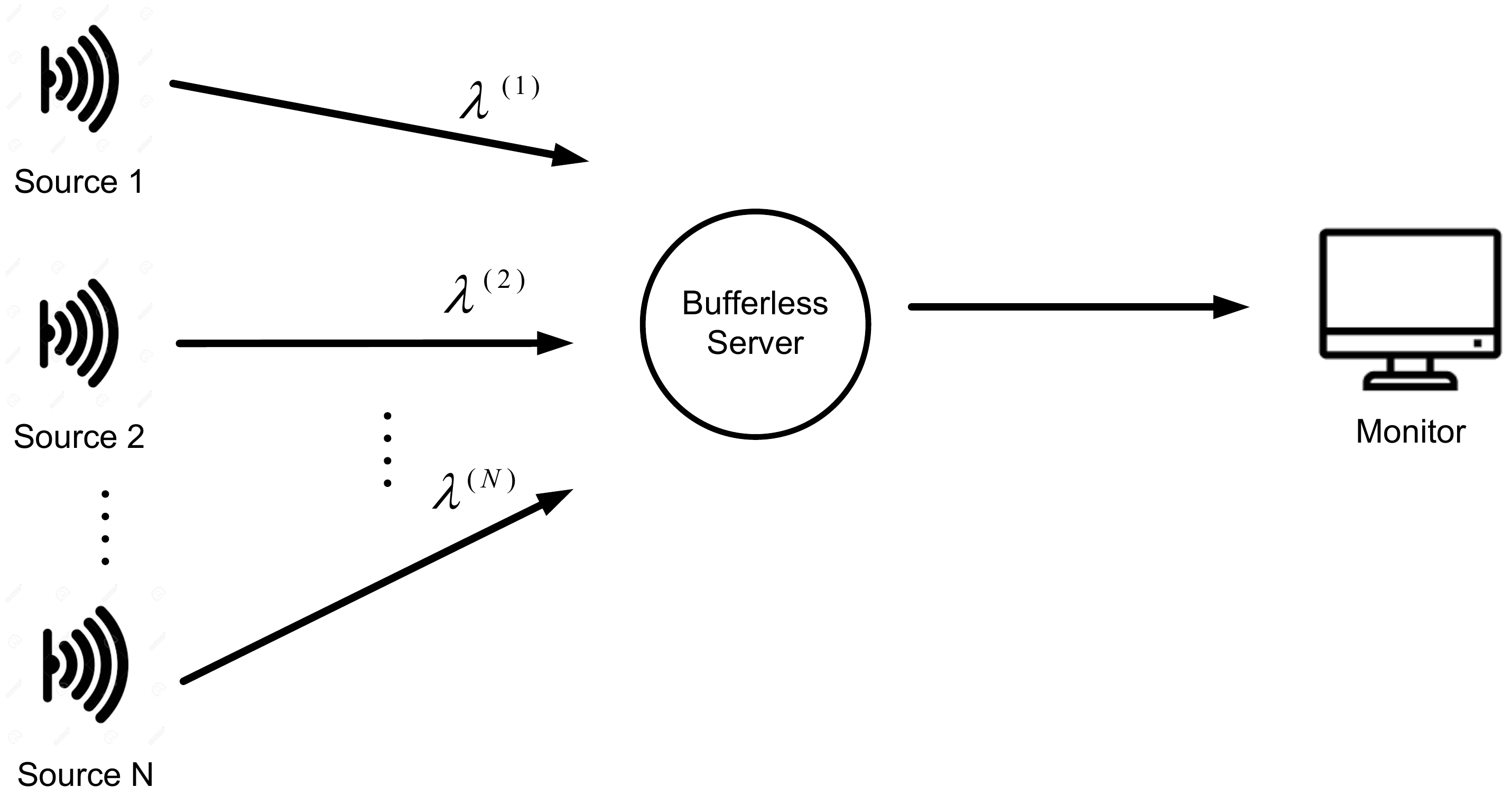}
	\caption{$N$ information sources sending status update messages through a bufferless server to a remote monitor.} 
	\label{fig:multisource}
\end{figure}

As the specific scenario, we consider the information update system in Fig.~\ref{fig:multisource} consisting of $N$ information sources each equipped with a sensor, a bufferless server local to the sources, and a remote monitor (or destination). The state of the source-$n$, $n=1,\ldots,N$ changes in time which is detected by its sensor and source-$n$ generates packets according to a Poisson process with intensity $\lambda_n$ that contain sensed data along with a time stamp, to be immediately forwarded to the server. 
Packets are sent by the server to the monitor via a communications network which introduces random delays, i.e., service time of packets, and the monitor immediately sends back positive acknowledgments to the server. In this paper, the server is assumed to be bufferless, i.e., no waiting room for information packets. A new arriving information packet immediately starts to receive service if the server is idle, but is either discarded or allowed to preempt the packet in service if the server is busy upon arrival. 
To make the model general, we assume the following:
\begin{itemize}
	\item  While the source-$n$ information packet is in service, a new information packet arrival from source $m,m=1,\ldots,N$ is allowed to preempt the packet in service with probability depending on both $n$ and $m$. The matrix composed of these preemption probabilities is called the preemption matrix of this system.
	\item The service time of each source has a PH-type distribution \cite{neuts81} and the requirements of sources are heterogeneous, i.e., each source has its own service time requirement.
	\item At the end of the service time needed for the delivery of an information packet, a packet error is said to occur with a probability depending on the source of this packet. This corrupt packet will be retransmitted with a per-source probability but is otherwise discarded. Therefore, when the server starts sending a packet, it stores a copy locally to potentially retransmit it. However, the system is still called bufferless since it does not provide a waiting room for new packets.
\end{itemize}
The above system is called the  multi-source (probabilistically) preemptive $M/PH/1/1$ queue with packet errors, which is the focus of this paper.  

\begin{figure*}[tb]
	\centering
	\includegraphics[width=\linewidth]{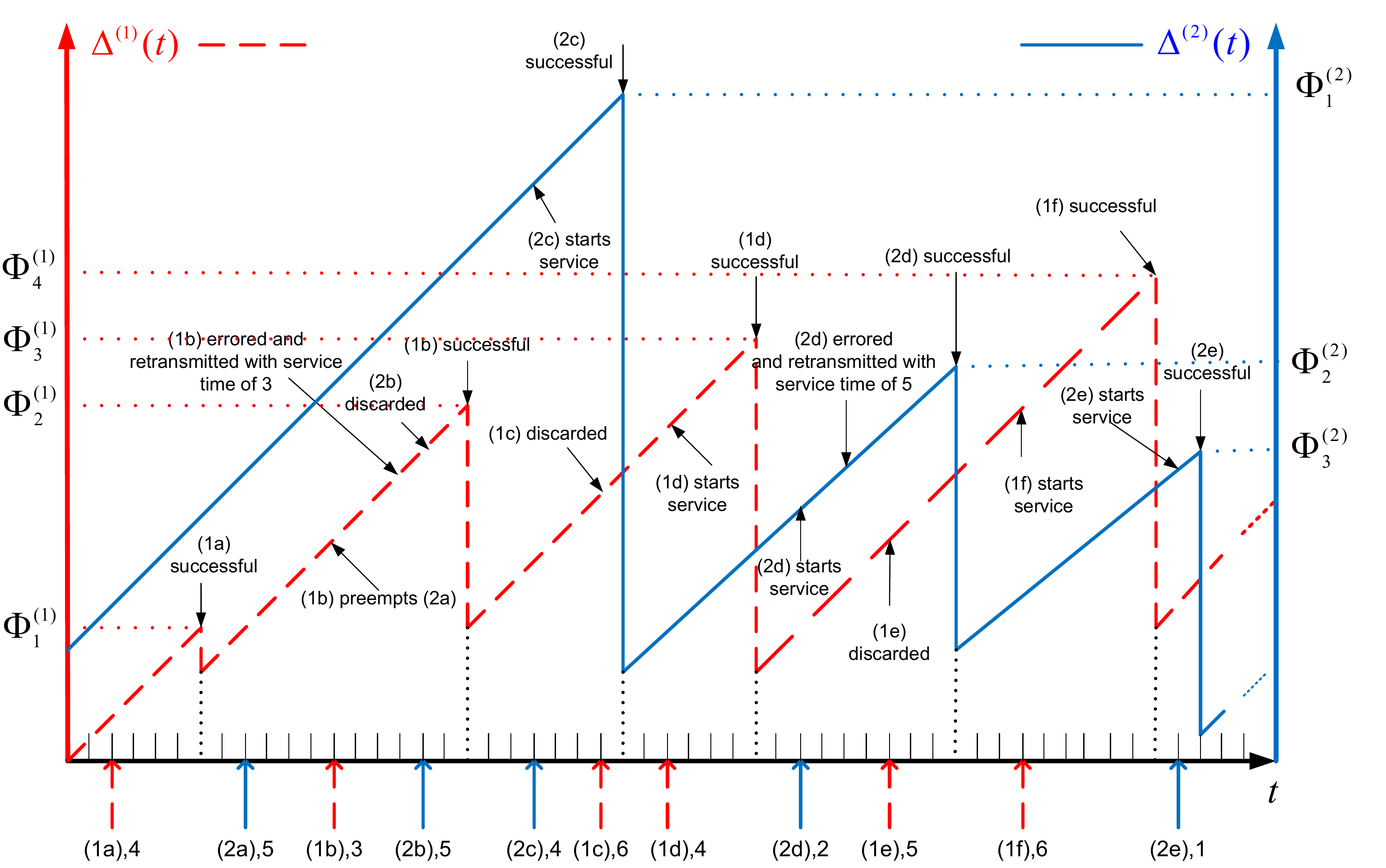}
	\caption{A sample path for the two AoI processes $\Delta^{(1)}(t)$ and $\Delta^{(2)}(t)$ in a 2-user bufferless system.} 
	\label{fig:samplepath1}
\end{figure*}

The AoI and PAoI processes will now be described for the bufferless server of interest followed by an illustrative example. 
First, we define a successful information packet as one which receives service until the end of the service time without preemption and without a packet error whereas other information packets are deemed unsuccessful. 
Let $t^{(n)}_j$ denote the arrival instant of the $j^{\text{th}}, j \geq 1$ successful source-$n$ information packet arriving at the server and let $\delta^{(n)}_j,j \geq 1$ denote the reception time at the monitor of the $j^{\text{th}}$ successful packet belonging to source $n$.
We denote by $\Delta^{(n)}(t), t\geq 0,$ the continuous-time random process with left-continuous sample paths representing the AoI for source-$n$ at time $t$ with a given initial condition $\Delta^{(n)}(0)$. At $t=0$, $\Delta^{(n)}(t)$ starts to increase linearly in time with a unit slope until the first successful packet reception at $t=\delta^{(n)}_1$.
The right limit $\Delta^{(n)}(\delta_1^+)=\lim_{t \downarrow \delta^{(n)}_1} \Delta^{(n)}(t)$ is set to $D^{(n)}_1$ where $D^{(n)}_j = \delta^{(n)}_j - t^{(n)}_j$  is the time spent  in service by the $j^{\text{th}}$ successful source-$n$ information packet. Subsequently, the process $\Delta^{(n)}(t)$ increases with unit slope until the next successful class-$n$ packet reception and the pattern repeats forever. Let $\Phi^{(n)}_j = \Delta^{(n)}(\delta_j), j \geq 1,$ denote the PAoI process
for source-$n$ which is a discrete-time continuous-valued random process associated with the AoI just at the epoch of packet receptions. 
Fig.~\ref{fig:samplepath1} illustrates the two sample paths of the random processes $\Delta^{(n)}(t),n=1,2$ in a two-source probabilistically preemptive bufferless server system with packet errors with the initial conditions $\Delta^{(1)}(0)=0$ and $\Delta^{(2)}(0)=5$.
The arrival epochs of the information packets are denoted by arrows at the bottom. The arriving packets are indexed as $a,b,\ldots$ with the notation ($n$$a$),$\theta$ indicating the first arrival from source-$n$ with a service time requirement of $\theta$. Let us now study Fig.~\ref{fig:samplepath1}. The first packet $(1a)$ from source-$1$ arrives at $t=2$ with a service time of 4. During its service time, no other arrivals take place and at the end of the service time, the packet $(1a)$ is received at $t=6$ without transmission errors. Therefore, in the time interval from $t=0$ to $t=6$, $\Delta^{(1)}(t)$ rises from the value 0 to $\Delta^{(1)}(6)=\Phi^{(1)}_1=6$ and $\Delta^{(1)}(6^+)$ is set to 4 which is the age of this packet, i.e., current time minus the time stamp on this packet. At $t=8$, packet $(2a)$
arrives and immediately starts to receive service with a service time requirement of 5, but the packet $(1b)$ arriving at $t=12$ preempts the service of $(2a)$ with a service time of 3. At $t=15$, this packet is received with packet error and is retransmitted with a service time of 3. At $t=16$, the new arrival $(2b)$ is discarded. At $t=18$, the packet $(1b)$ is successfully received. In the time interval from $t=6$ to $t=18$, the AoI process $\Delta^{(1)}(t)$ rises from the value 4 to $\Phi^{(1)}_2=16$ after which $\Delta^{(1)}(16^+)$ is set to 6 which is the sum of two service times 3 and 3 required for transmission of the successfully received packet 
$(2b)$. For source-$2$, the first successfully received packet is $(2c)$ which arrives at $t=19$ with a service time requirement of 4 and is not preempted by the packet $(1c)$ and is successfully received at $t=23$ without error. Therefore, $\Delta^{(2)}(t)$ rises from its initial value 5 to $\Phi^{(2)}_1=28$ after which $\Delta^{(2)}(23^+)$ is set to the age of this particular packet at $t=23$. These patterns repeat for both these processes.

We use the notation $\Delta^{(n)}$, $\Phi^{(n)}$, and $D^{(n)}$ to denote the steady-state random variables associated with the processes $\Delta^{(n)}(t)$, $\Phi_j^{(n)}$, and $D_j^{(n)}$, respectively. For the general $N$-source probabilistically preemptive $M/PH/1/1$ system with packet errors (an illustration of which is given in Fig.~\ref{fig:samplepath1} for $N=2$), we are interested in finding the following steady-state cdfs (cumulative distribution function) for the random variables $\Delta^{(n)}$ and $\Phi^{(n)}$, respectively:
\begin{align}	
F_{\Delta^{(n)}}(x) &= \lim\limits_{t\to \infty }   \Pr \{ \Delta^{(n)}(t) \leq x \}, \ x \geq 0, \\
F_{\Phi^{(n)}}(x) &= \lim\limits_{j \to \infty }   \Pr \{ \Phi^{(n)}_j \leq x \}, \ x \geq 0.
\label{cdf}
\end{align}
Note that $F_{\Delta^{(n)}}(0)$ and $F_{\Phi^{(n)}}(0)$ must be zero since there can not be a probability mass at the origin for these two processes.
Also, let $f_{\Delta^{(n)}}(x)$ and $f_{\Phi^{(n)}}(x)$ for $x \geq 0$ denote the corresponding steady-state pdfs (probability distribution function) with the corresponding non-central moments:
\begin{align}
E\left[(\Delta^{(n)})^i \right]&=\int_{0}^{\infty} x^i f_{\Delta^{(n)}}(x) dx, \\
E\left[ (\Phi^{(n)})^i \right]&=\int_{0}^{\infty} x^i f_{\Phi^{(n)}}(x) dx.
\end{align}

We observe from Fig.~\ref{fig:samplepath1} that each cycle of the AoI process $\Delta^{(n)}(t)$ consists of a linear curve with unit slope that starts at value $D^{(n)}_j$ for some successful packet index $j$ for source-$n$ and terminates at the peak value $\Phi^{(n)}_{j+1}$. Subsequently, the sample path of the AoI process $\Delta^{(n)}(t)$ consists of an ordered concatenation of infinitely many cycles each of which behaves as described above.  For obtaining the distribution of the AoI and PAoI processes, we propose to use the theory of Markov Fluid Queues (MFQ) \cite{anick_mitra82,kosten.1984,kulkarni_1997}. Existing steady-state MFQ solvers that we propose to use are matrix analytical and they rely on numerically stable and efficient vector-matrix operations. The main idea is that we construct MFQs that produce sample paths whose certain parts  coincide with the sample cycles  of the AoI process. Additionally, these MFQ-produced cycles contain sample values that coincide with the sample values of the PAoI process. Hence, the exact distributions of the AoI and PAoI processes given in \eqref{cdf} can be obtained out of the steady-state solution of certain MFQs, the construction of which is the main focus of this paper. The main contribution of this paper is to obtain the exact distribution of AoI and PAoI processes in a very general $M/PH/1/1$ framework representative of bufferless servers arising in information status update systems using the theory of MFQs. 

The main contributions of this paper are the following:
\begin{itemize}
	\item  We present an analytical model for bufferless servers arising in status update systems that can be globally preemptive, self-preemptive, non-preemptive, etc. with a unifying probabilistic preemption framework. Besides unification, probabilistic preemption can be optimum depending on how rewards or costs are defined.
	\item Packet errors and heterogenous service time requirements across sources makes the model even more general than the existing ones.
	\item Most existing results use SHS and cope with less general settings while providing  means to obtain the average AoI and PAoI values while falling short in most of the cases in obtaining their exact distributions (see subsection~\ref{ms}). In this paper, we obtain the exact distributions of the AoI and PAoI processes numerically which can be crucial for system design.	
\end{itemize}

The organization of the paper is as follows. In Section~2, related work is presented. Section~3 presents the notation throughout the paper as well as preliminaries on PH-type distributions and MFQs. Section~4 presents the analytical model. In Section~5, we provide numerical examples to validate the proposed approach as well as examples to study the impact of preemption probabilities on system performance with respect to certain AoI- and PAoI-related performance metrics. Finally, we conclude in Section~6.

\section{Related Work}
The AoI concept was first introduced in \cite{kaul_etal_infocom12} in the context of a single-source, single-server $M/M/1$ queueing model. 
This model is then extended to multiple sources in \cite{yates_kaul_ISIT12} and since then the single-server model in Fig.~\ref{fig:multisource} for AoI has extensively been used for status update systems in the literature \cite{kaul_etal_infocom12,kosta_etal_survey}. There are many variations of this single-server queueing system studied in the recent literature depending on
\begin{itemize}
	\item Whether there is a single-source or multiple sources, feeding the queue with information packets,
	\item Whether there is a transmission error or not,
	\item Whether the interest in on the mean AoI and PAoI values, or their exact distributions are sought,
	\item Generality of the distributions assumed for the interarrival times and service times,
	\item Queue capacity which represents the maximum number of packets that are allowed to be in the system, including those in the waiting room and the one in service,  
	\item Scheduling discipline to be used whether it be First Come First Serve (FCFS), Preemptive Last Come First Serve (P-LCFS), Non-preemptive LCFS (NP-LCFS), etc. 
	\item Use of buffer management schemes in charge of packet dropping at the server.
\end{itemize}
The related work on AoI analytical models is summarized below first in the single-source setting. Subsequently, related work will be presented regarding the existing multi-source queueing models.
\subsection{Single-source Queueing Models}	
\label{ss}
In \cite{kaul_etal_infocom12}, the mean AoI is obtained for the single-source $M/M/1$, $M/D/1$, and $D/M/1$ queues with infinite buffer capacity and FCFS scheduling. 
In \cite{inoue_etal_ISIT17}, expressions are derived
for the LST transform of the stationary distributions
of the AoI and the PAoI processes in $M/GI/1$ and $GI/M/1$ queues. 
Despite the fact that relatively large buffers and FCFS scheduling are the de-facto choices in operational packet-switched communication networks,
such choices have been shown to give rise to poor AoI performance in moderate to high load regimes.
The reference \cite{costa_etal_TIT16} studies the AoI and PAoI distributions for small buffer systems, including the conventional $M/M/1/1$ and $M/M/1/2$ queues, as well as the so-called $M/M/1/2^{\ast}$ queue, for which the packet waiting in the queue is to be replaced by a newer packet arrival, 
which actually is a non-preemptive LCFS system. The mean AoI and PAoI figures in the pre-emptive LCFS $M/G/1/1$ queueing system is studied in \cite{najm_nasser_isit16}
where a new arrival preempts the packet in service and the service time distribution is assumed to follow a more general gamma distribution.
Exact PAoI expressions are derived in an $M/M/1$ queueing system with packet delivery errors using different scheduling policies such as FCFS, P-LCFS, and NP-LCFS in \cite{chen_huang_isit16}.
Exact expressions for the stationary distributions of AoI and PAoI for a very wide class of single-source information update systems are given in \cite{inoue_etal_tit19}.
A recent work in \cite{akar_etal_tcom20} also obtains the exact distributions of AoI and PAoI in bufferless systems with probabilistic preemption and PH-type distributions for both interarrival and service times. A similar model is also proposed in \cite{akar_etal_tcom20} for a single-buffer queueing system with Poisson packet arrivals and PH-distributed service times allowing probabilistic replacement of the waiting packet by a newer packet arrival.
A discrete-time queueing model with Bernoulli arrivals and geometric service times, using FCFS and non-preemptive LCFS scheduling is presented in \cite{kosta_etal_isit19} with expressions for the mean AoI and PAoI values.
In addition to exact methods, a number of studies provide bounds for certain AoI-related metrics of interest.
The reference \cite{soysal_ulukus_unpublished} derives upper bounds for the mean AoI for
the $G/G/1/1$ queue as well as its preemptive version while showing that the bounds are close to actual values. Similarly, the authors of \cite{champati_etal_infocom19}  present a method for obtaining upper bounds for the AoI violation probability for both $GI/GI/1/1$ and $GI/GI/1/2^{\ast}$ systems, in addition to some exact closed-form expressions for some sub-cases.
\subsection{Multi-source Queueing Models}
\label{ms}
The reference \cite{huang_modiano} derives the mean PAoI expression for $M/G/1$ and $M/G/1/1$ systems with heterogeneous service time requirements which enables one to optimize system cost, as a function of mean PAoI, by choice of the update interval. The authors of \cite{yates_kaul_tit19} study the multi-source $M/M/1$ model with FCFS as well as two variations (preemptive and nonpreemptive with replacement) of LCFS using the theory of SHS and obtain exact expressions for the mean AoI.
A preemptive $M/G/1/1$ queue is considered in \cite{najm2018status} with a common service time for all sources in which expressions for the mean AoI and PAoI are derived.
A similar preemptive $M/G/1/1$ system is studied in \cite{farazi_etal_Asilomar19} allowing packet delivery errors. The authors \cite{farazi_etal_Asilomar19} allow preemption of a source in service by a newly-arriving packet from the same source and derive the mean AoI expressions for each source using SHS technique.
The reference \cite{moltafet2020average} considers a two-source $M/M/1/2$ queueing system in which a packet waiting in the queue can be replaced only by a newly-arriving packet from the same source, again using SHS techniques. A non-preemptive $M/M/1/m$ with common service times across sources is again studied by the SHS technique in \cite{kaul2020timely} and mean AoI expressions are derived. A more general hyperexponential ($H_2$) service time distribution for each class is considered in \cite{yates_etal_isit19} for an $M/H_2/1/1$ nonpreemptive bufferless queue to derive an expression for the mean AoI per class.    

\section{Preliminaries}
Uppercase bold letters are used to denote real-valued matrices. Lowercase bold (plain) letters or symbols are used to denote real-valued vectors (scalars).
The $(i,j)^{\text{th}}$ th entry of $\bm{A}$ is denoted by ${A_{i,j}}$ and the $j^{\text{th}}$ entry 
of a row or column vector $\bm{\alpha}$ is ${\alpha}_j$. 
The notations $\bm{0}_{k \times \ell} $, ${\bm I_m}$, and ${\bm 1_n}$ are used to denote the matrix of zeros of size $k \times l$, identity matrix of size $m$, and a column matrix of
ones of size $n$, respectively. When used without a subscript, it is left to the reader to infer the size information from the context.  
Let $\bm{A}$ be an $n \times m$ matrix and $\bm{B}$ a $p\times q$ matrix. The Kronecker product of the matrices $\bm{A}$ and $\bm{B}$ 
is denoted by  $\bm{A} \otimes \bm{B}$ which is of size $np \times mq$.
The notation $\textbf{diag}\{\bm{A},\bm{B}\}$ denotes the block diagonal concatenation of the matrices $\bm{A}$ and $\bm{B}$ and is diagonal if the individual matrices $\bm{A}$ 
and $\bm{B}$ are diagonal.
%A matrix $A$ is quasi-upper triangular is upper triangular with the exception that it may have either 1-by-1 or 2-by-2 blocks on its diagonal corresponding to real and complex eigenvalues of the matrix $A$, respectively. 
A square matrix is said to be stable (anti-stable) if each of its eigenvalues has negative (non-negative) real parts. 
The notation $\left[\bm{\alpha},\bm{\beta}  \right]$ is used for the concatenation of the two row vectors $\bm{\alpha}$ and $\bm{\beta}$. 
The function $u(x)$ refers to the Heaviside step function, also known as the unit step function, whereas $\delta(x)$ stands for the Dirac delta function, also known as the unit impulse function.
\subsection{Phase-type Distributions}
\label{ph} 
In the context of queueing systems, Phase-type (PH-type) distributions are often used for modeling independent and identically distributed (iid)
non-exponential interarrival and/or service times \cite{neuts81}.  Using PH-type distributions gives rise to 
algorithmically tractable methods for finding both the steady-state and transient solutions of such queueing systems; see \cite{asmussen_etal_SJS96} and the references therein. 
For rigorous description of PH-type distributions, we first define a Markov process  on the state-space
$\mathcal{S} = \{1,2,\ldots,m,m+1\}$ with $m$ transient states, one absorbing state $m+1$, initial probability vector $\left[ \bm{\sigma},\sigma_0\right] $, and an
infinitesimal generator of the form 
\[ \left[ \begin{array}{{c;{2pt/2pt}c}}
\bm{S} & \bm{\nu} \\ \hdashline[2pt/2pt]
\bm{0} & 0 \end{array}
\right],
\]
where $\bm{\sigma}$ is a row vector of size $m$, $\sigma_0=1-\bm{\sigma} {\bm  1}$  is a scalar, the sub-generator
$\bm{S}$ is $m \times m$, and $\bm{\nu}$ is a column vector of size $m$ such that $\bm{\nu}=-\bm{S}{\bm 1}$. The time
to absorption to the absorbing state $m+1$, say $X$, is said to be PH-type characterized with the pair $(\bm{\sigma},\bm{S})$, i.e., $X \sim PH(\bm{\sigma},\bm{S})$.
In most typical scenarios, $\sigma_0$ is zero. 
%where the notation $\sim$ is synonymous with \enquote{distributed according to}.
%Note that $\left[ \bm{\sigma}, \sigma_0 \right]$ is a probability vector 
%and the diagonal elements of $S$ are strictly negative, its off-diagonal elements are non-negative and $S {\bm 1}\leq \bm{0}$ elementwise. 
The cdf and the pdf of $X \sim PH(\bm{\sigma},\bm{S})$, denoted by $F_X(x)$ and $f_X(x)$, respectively, are given as: 
\begin{equation}
F_X(x) =(1 -\bm{\sigma} e^{Sx} {\bm 1}) \:  u(x), \;
f_X(x) =-\bm{\sigma} e^{Sx} S {\bm 1} \:  u(x) + \sigma_0 \:  \delta(x).
\label{phdensity}
\end{equation}
PH-type distributions are dense in the field of all positive-valued distributions and therefore they can principally be used to approximate any positive-valued distribution \cite{ocinneide}. 
Given sample data or an arbitrary pdf, one can use one of the existing algorithms, such as the Expectation Maximization (EM) algorithm of \cite{asmussen_etal_SJS96} for maximum likelihood estimation, or the moment-matching algorithm of \cite{PhFit}, or the statistical inference-based algorithm of \cite{HiroyukiOkamura2016}, to construct a PH-type distribution that matches data or accurately approximates the given pdf. 
\subsection{Markov Fluid Queues}
\label{mfq}
We describe an MFQ by a joint Markovian process  
%\begin{equation*}
%\begin{aligned}
${\bm X(t)}= (X_f(t),X_m(t))$, $t\geq 0$, where
$0\leq X_f(t) < \infty$ is the continuous-valued fluid level in the buffer and $X_m(t) \in {\mathcal S}= \{1,2,\ldots,n\}$
is the modulating phase process $X_m(t)$ which behaves as a Continuous Time Markov Chain (CTMC) with state space ${\mathcal S}$ and generator $\bm{Q}$ $(\bm{\tilde{Q}})$ when $X_f(t) > 0$ ($X_f(t)=0$).  The parameter $n$ is the system size. 
%Finite MFQs in which the fluid level is not allowed to exceed a certain finite level are outside the scope of this paper. 
The drift (rate of fluid change) of the MFQ equals $r_i$ when the modulating process $X_m(t)$ visits state $i$ and $\bm{R}$ is defined as the diagonal matrix of drifts: 
$ \bm{R}=\textbf{diag}\{
r_1, r_2, \ldots, r_{n} \}$. When $X_f(t)=0$ and $X_m(t)=i$ with $r_i < 0$, $X_f(t)$ sticks to the boundary at zero.  The process ${\bm X(t)}$ is said to be characterized with the matrix triple $(\bm{Q},\bm{\tilde{Q}},\bm{R})$, i.e., ${\bm X(t)} \sim MFQ(\bm{Q},\bm{\tilde{Q}},\bm{R})$.
In most existing studies, $\bm{Q} = \bm{\tilde{Q}}$, for which stationary solutions are derived by \cite{kulkarni_1997},\cite{anick_mitra82} by using the eigendecomposition of a  certain matrix and also by \cite{akar_sohraby_jap04} using the matrix sign function avoiding the computation of the eigenvectors, a problem known to be ill-conditioned \cite{golub.vanloan.1996}.
The more general $\bm{Q} \neq \bm{\tilde{Q}}$ case turns out to be a sub-case of multi-regime MFQs whose steady-state solutions can be obtained through the ordered Schur decomposition, again avoiding ill-conditioned eigendecompositions \cite{kankaya.2008}.  For other eigendecomposition-free numerically efficient and stable algorithms for solving multi-regime MFQs, we refer the reader to the matrix-analytical approaches of \cite{soares_latouche} and \cite{horvath_vanhoudt}.

We assume $r_i \neq 0,\ 1 \leq i \leq n$ and 
$r_i > 0,\ i \leq b$ and $r_i < 0,\ i > b$, since otherwise states can always be reordered for this purpose.
We are interested in finding the steady-state joint pdf vector
\begin{align}
\bm{f(x)}  &=  \left[  f_1(x), f_2(x),\ldots,f_{n}(x) \right], \\
f_i(x)  &=  \lim\limits_{t\to \infty } \frac{d}{dx}  \Pr\{X_f(t)\leq x, X_m(t)=i\}, \; x >0 ,\label{density}
\end{align}
and the steady-state probability mass accumulation (pma) vector at zero:
\begin{equation}
\bm{c} = \left[  c_1, c_2, \ldots, c_{n}  \right], \quad
c_i  =  \lim\limits_{t\to \infty } \Pr \{X_f(t)=0, X_m(t)=i\}. \label{accumulation}
\end{equation}
We now describe the method of \cite{kankaya.2008} without proof, in three steps, adapted to MFQs described above, to find the quantities of interest in \eqref{density} and \eqref{accumulation}, when they exist. 
In Step 1, we find an orthogonal matrix $\bm{U}$ such that the following holds: 
\begin{equation}
\bm{U}^T \bm{Q} \bm{R}^{-1} \bm{U} =\left[ \begin{array}{c;{2pt/2pt}c}
\bm{\Psi}_{a \times a} & \ast  \\ \hdashline[2pt/2pt]
\bm{0} & \bm{A}_{b \times b}
\end{array} \right], \quad \bm{U}^T = \left[ \begin{array}{c} 
\ast \\ \hdashline[2pt/2pt]
\bm{H}_{b \times n}
\end{array} \right],  \label{step1}
\end{equation}
for an anti-stable matrix $\bm{\Psi}$ with an eigenvalue at the origin, stable matrix $\bm{A}$, and $\ast$ denoting an arbitrary sub-matrix. The well-known ordered real Schur form (available in Lapack, Matlab, and Octave software packages) can be used towards obtaining the decomposition \eqref{step2} \cite{golub.vanloan.1996}. 
In Step 2, we solve for the $1 \times b$ vector $\bm{g}$ and $1 \times a$ vector $\bm{d}$ from the following linear matrix equation:
\begin{equation}
\left[ \begin{array}{c;{2pt/2pt}c}
\bm{g} & \bm{d}
\end{array} \right]
\left[ \begin{array}{c;{2pt/2pt}c}
\bm{H}\bm{R} & -\bm{A}^{-1}\bm{H}{\bf 1}_n \\ \hdashline[2pt/2pt]
-\bm{\tilde{Q}}^{\ast} & {\bf 1}_a 
\end{array} \right] = 
\left[ \begin{array}{c;{2pt/2pt}c}
{\bm 0}_{1 \times n} & 
1
\end{array} \right], \label{step2}
\end{equation}
with $\bm{\tilde{Q}}^{\ast}$ denoting the matrix composed of the last $a$ rows of $\bm{\tilde{Q}}$. 
Finally, in Step 3, we write
\begin{equation}	
\bm{f(x)} = \bm{g} e^{\bm{A}x} \bm{H} \ u(x) + \bm{c} \ \delta(x),  \ f_i(x)= \bm{g} e^{\bm{A}x} {\bm h}_i \ u(x)+ c_i \ \delta(x),  \label{meform}
\end{equation}
where ${\bm h}_i$ denotes the $i^{\text{th}}$ column of $\bm{H}$.		
%The computational complexity of the first two steps are $\mathcal{O}(n^3)$ in Alg.~1.
%In an important sub-case that will be shown to arise in AoI queueing models later in this paper, a simple explicit way of complexity $\mathcal{O}(n^2)$ to find a matrix $P$ in \eqref{step2} of Alg.~1 is presented in the following lemma based on the Householder transformation given in \cite{golub.vanloan.1996}.
%\begin{lemma}
%	\label{householder}
%	Consider the process ${\bm X(t)} \sim GMFQ(Q,\tilde{Q},R)$ with order $n$, $R = \textbf{diag} \{ \bm{I},-1\}$, and the last row of $Q$ is the zero matrix. Let $\bm{u_1}$ be a column vector of ones of size $n$ except for the last entry which is minus one. Also, let $\bm{u_2}$ be a column vector of zeros except for the first entry which is one. 
%	Let $u=\bm{u_1} - || \bm{u_1} ||_2 \bm{u_2}$. 
%	Then, the symmetric orthogonal matrix $P$ defined by $P={\bm I} - \frac{2 \bm{uu^T}}{\bm{u^T u}}$ gives rise to the factorization \eqref{step2} with the scalar $F$ being zero. \end{lemma}
\section{Analytical Model for the Multi-source Preemptive $M/PH/1/1$ Queue}
We consider the status update system in Fig.~\ref{fig:multisource} with $N$ sources, a server, and a monitor. The source $n, n=1,\ldots,N$ generates packets that carry status update information, according to a Poisson process with intensity $\lambda_n$. 
The traffic intensity vector is denoted by $\bm{\lambda} = (\lambda_1,\ldots,\lambda_N)$. 
We define the total arrival rate $\lambda = \sum_{n=1}^N \lambda_n$. A source-$n$ information packet immediately starts to receive service from the server when it finds the server idle upon arrival and its service time $\Theta^{(n)} \sim PH(\bm{\sigma}^{(n)},\bm{S}^{(n)})$ with order $\ell_n$, $\bm{\nu}^{(n)}=-\bm{S}^{(n)} \bm{1}$, and $E[\Theta^{(n)}] = \frac{1}{\mu_n}$. Let the per-source load be defined as $\rho_n = \frac{\lambda_n}{\mu_n}$ and the total load $\rho = \sum_{n=1}^N \rho_n$.
We define the total order $\ell_T = \sum_{n=1}^N \ell_n$.
We are given a preemption matrix $\bm{P}$ so that while the source-$n$ information packet is in service, a new information packet arrival from source $m,m=1,\ldots,N$ is allowed to preempt the packet in service with probability ${P}_{n,m}$. The following sub-cases of a general preemption matrix $\bm{P}$ have been studied in the literature:
\begin{itemize}
	\item $\bm{P}=\bm{0}$ refers to a non-preemptive system \cite{kosta_etal_survey},
	\item $\bm{P}=\bm{1_N} \bm{1_N^T}$ case is referred to as {\em global preemption} in \cite{yates_kaul_tit19},
	\item $\bm{P}=\bm{I}$ case is referred to as {\em self preemption} in \cite{farazi_etal_Asilomar19}.
\end{itemize}

For each source $n$, we define the total intensity of traffic that can preempt a source-$n$ packet in service as $\bar{\lambda}_n = \sum_{m=1}^N \lambda_m {P}_{n,m}$. 
In these systems, preemption may potentially be beneficial for two different purposes: (i) a new information packet always carries more timely information than the one in service, (ii) sources can be differentiated from each other by proper choice of preemption probabilities. At the end of the service time of a source-$n$ packet, a transmission error is detected at the monitor with probability $e_n$. We also denote the successful transmission probability of a source-$n$
packet by $q_n = 1- e_n$.
An errored class-$n$ packet (irrespective of how many times it was transmitted) is retransmitted with probability $r_n$ whereas it will be discarded with probability $d_n = 1- r_n$. Recall that an information packet which receives service until the end of the service time without preemption and without a packet error is called a successful packet. 

We tag a specific source, say source 1, for which the exact distributions of the AoI and PAoI processes are to be obtained. 
%For the sake of convenience, the steady-state AoI 
%process $\Delta(t) :=\Delta^{(1)}(t)$, the steady-state PAoI process $\Phi = \Phi^{(1)}$, and the system time of successful packets $D = D^{(1)}$ respectively. 
If the interest is on another information source-$n$ where $n \neq 1$, the same procedure can be repeated by renumbering the sources.
For this purpose, we construct an MFQ process ${\bm X(t)}=(X_f(t),X_m(t))$ by which we have a single fluid level trajectory of infinitely many cycles where each cycle comprises four stages, namely stages 1-4, that are described as follows. 
Every cycle begins with stage 1 at which the service of a source-$1$ packet begins. If this packet is preempted or errored, we go back to stage 1 through stage 4.
When a source-$1$ packet is eventually received successfully, we transition from stage 1 to stage 2 while ensuring that the fluid level at this transition epoch is distributed according to $D^{(1)}$ which actually is the system time of successful source-$1$ packets. In stage 2, we wait for the next packet arrival from any one of the sources upon which we transition to stage 3. 
In stage 3, we are within the service time of a source-$n$ packet for some source $n$. If a source-$1$ service is successfully over, we end the cycle by transitioning to stage 1 again through stage 4.  If a source-$n$ service for $n \neq 1$ is successfully over during stage 3, we go back to stage 2 waiting for a new packet arrival. 
In the case of preemption or transmission error followed by retransmission during stage 3, we stay at stage 3.  For the case of error for a source-$n$ packet with discarding, a transition to stage 2 occurs. More formally,
\begin{itemize}
	\item During stage 1, the modulating process $X_m(t)$ visits states $j$, $ 1 \leq j \leq \ell_1$ that keep track of the phase of the service time of the source-$1$ packet.
	\item Stage 2 consists of one single state $0$ waiting for an information packet arrival from one of the information sources. 
	\item During stage 3, $X_m(t)$ visits the states $(i,j)$, where $1 \leq i \leq N$ keeps track of the source index of the packet in service and $j, 1 \leq j \leq \ell_i$ keeps track of the phase of the service time of the packet in service. 
	\item Stage 4 consists of one single final state $-1$ by which we prepare for starting the next cycle. 
\end{itemize} 
%During stage 1, the modulating process $X_m(t)$ visits states $j$, $ 1 \leq j \leq \ell^{(1)}$ that keep track of the phase of the service time of the source-$1$ packet. On the other hand, stage 2 consists of one single state $0$ waiting for an information packet arrival from one of the information sources. During stage 3, $X_m(t)$ visits the states $(i,j)$, where $1 \leq i \leq N$ keeps track of the source index of the packet in service and $j, 1 \leq j \leq \ell^{(i)}$ keeps track of the phase of the service time of the packet in service. Finally, stage 4 consists of one single final state $-1$ by which we prepare for the next cycle. 
Let us now describe the operation of the MFQ ${\bm X(t)}$. A cycle of ${\bm X(t)}$ begins with a visit to a state in stage 1 when the fluid level is zero. During stage 1, the fluid level rises with a unit slope. Let us assume that we are at state $j$, $ 1 \leq j \leq \ell_1$ in stage 1 during which the arrival processes of all sources are turned on. There are five possible transitions:
\begin{itemize}
	\item with rate ${S}^{(1)}_{j,k}$, a transition to state $k$ occurs, 
	\item with rate $\bar{\lambda}_1 = \sum_{n=1}^N {P}_{1,n} \lambda_n$, the source-$1$ packet is preempted by a new arrival; and a transition to state $-1$ occurs,
	\item with rate ${\nu}^{(1)}_j e_1 d_1$,  the service time of the packet is over but the packet is errored and discarded; and a transition to state $-1$ occurs,	
	\item with rate ${\nu}^{(1)}_j e_1 r^{(1)} {\sigma}^{(1)}_k $,  the service time of the packet is over and it is errored and retransmitted; and a transition to state $k$ occurs,	
	\item with rate ${\nu}^{(1)}_j q_1$,  the service time is over and the packet is successful and a transition to state $0$ occurs.
\end{itemize}
With the transitions described above, we ensure that when we are at the beginning of state $0$, the fluid level has risen to a level distributed according to $D^{(1)}$. During state $0$, the fluid level rises with unit slope and with rate $\lambda^{(i)} {\sigma}^{(i)}_j$, a transition to state $(i,j)$ occurs in stage 3. 

Let us now assume that we are at state $(i,j)$ in stage 3 during which the fluid level continues to rise again with unit slope. Therefore, a source-$i$ packet is in service and we are in phase $j$ of its service time.  We have the following transition possibilities from state $(i,j)$:
\begin{itemize}
	\item with rate ${S}^{(i)}_{j,k}$, a transition to state $(i,k)$ occurs,
	\item with rate  ${P}_{i,n} \lambda_n$, the source-$i$ packet is preempted by a new information packet from source $n$ and a transition to state $(n,l)$ occurs with rate  ${P}_{i,n} \lambda_n {\sigma}^{(n)}_l$,
	\item with rate ${\nu}^{(i)}_j e_i d_i$,  the service time of the packet is over but the packet is errored and discarded; and a transition to state $0$ occurs,	
	\item with rate ${\nu}^{(i)}_j e_i r_i {\sigma}^{(i)}_k $,  the service time of the packet is over and it is errored and retransmitted; and a transition to state $(i,k)$ occurs,	
	\item when $i \neq 1$, with rate ${\nu}^{(i)}_j q_i$,  the service time of the source-$i$ packet is over and is successful; a transition to state $0$ occurs,
	\item  when $i=1$, with  rate ${\nu}^{(1)}_j q_1$,  the service time of the source-$1$ packet is over and is successful giving rise to a transition to state $-1$.	    
\end{itemize}
When at state $-1$, the fluid level always drops with a rate of minus one without any state changes until the fluid level zero is hit. The arrival process is turned off in this stage and the only way to escape from this particular state is through a transition to state $i$ with transition rate 
${\sigma}^{(1)}_i$.
With the lexicographical ordering of the states from stages 1 to 4,
${\bm X(t)} \sim MFQ(\bm{Q},\bm{\tilde{Q}},\bm{R})$ with system size $\ell_T + \ell_1 + 2$ where $\bm{Q}$ equals to the matrix given in Eqn.~\eqref{MatrixQ}

\begin{equation}
	{
		Q=\left[
		\begin{array}{c;{2pt/2pt}c;{2pt/2pt}c;{2pt/2pt}c;{2pt/2pt}c;{2pt/2pt}c;{2pt/2pt}c}
		\begin{array}{c}
		\bm{S}^{(1)} - \bar{\lambda}_1 \bm{I}\\
		+e_1  r_1 \bm{V}^{(1)} 
		\end{array} & q_1 \bm{\nu}^{(1)}  & \bm{0} & \bm{0} & \cdots & \bm{0} & 
		\begin{array}{c} \bar{\lambda}_1 \bm{1} \\
		+ e_1  d_1  \bm{\nu}^{(1)} \end{array}\\ \hdashline[2pt/2pt] 
		\bm{0} & -{\lambda} & \lambda_1 \bm{\sigma}^{(1)} & \lambda_2 \bm{\sigma}^{(2)} & \cdots &  \lambda_n \bm{\sigma}^{(N)} & 0 \\ \hdashline[2pt/2pt]
		\bm{0} & e_1 d_1\bm{\nu}^{(1)}  & \begin{array}{c} 
		\bm{S}^{(1)} - \bar{\lambda}_1 \bm{I}  \\ +\lambda_1 {P}_{1,1} \bm{F}^{(1)} \\ 
		+e_1  r_1 \bm{V}^{(1)} \
		
		\end{array}
		&  \lambda_2 {P}_{1,2} \bm{F}^{(2)} & \cdots &  \lambda_N {P}_{1,N} \bm{F}^{(N)} & q_1 \bm{\nu}^{(1)} \\ \hdashline[2pt/2pt]
		\bm{0} & \begin{array}{c} q_2 \bm{\nu}^{(2)} \\ 
		+ e_2 d_2 \bm{\nu}^{(2)} \end{array}& \lambda_1 {P}_{2,1} \bm{F}^{(1)} & 
		\begin{array}{c} \bm{S}^{(2)} - \bar{\lambda}_2 \bm{I} \\  +  \lambda_2 {P}_{2,2} \bm{F}^{(2)} \\ 
		+e_2  r_2 \bm{V}^{(2)} 
		\end{array} &
		\cdots &
		\lambda_N {P}_{2,N} \bm{F}^{(N)} &
		\bm{0} \\ \hdashline[2pt/2pt]
		\vdots & \vdots & \vdots & \vdots & \ddots & \vdots & \vdots  \\ \hdashline[2pt/2pt]
		\bm{0} & \begin{array}{c} q_N \bm{\nu}^{(N)} \\ 
		+ e_N d_N \bm{\nu}^{(N)} \end{array} & \lambda_1 {P}_{N,1} \bm{F}^{(1)} & 
		\lambda_2 {P}_{N,2} \bm{F}^{(2)}&
		\cdots &   \begin{array}{c} \bm{S}^{(N)} - \bar{\lambda}_N \bm{I}\\  + \lambda_N {P}_{N,N} \bm{F}^{(N)} \\
		+e_N  r_N \bm{V}^{(N)} \\
		\end{array}& \bm{0}  \\ \hdashline[2pt/2pt]
		\bm{0} & 0 & \bm{0} & \bm{0}  & \bm{0} &  0 & 0 
		\end{array}
		\right],
		\label{MatrixQ}
	}
\end{equation}

and 
\begin{equation}
\bm{V}^{(j)} = \bm{\nu}^{(j)} \otimes \bm{\sigma}^{(j)}, \quad \bm{F}^{(j)} = \bm{1} \otimes \bm{\sigma}^{(j)},
\end{equation}
the matrix $\bm{\tilde{Q}}$ is the same as $\bm{Q}$ except for the block entry in the south-west corner which is set to $\bm{\sigma}^{(1)}$,
and for the scalar at the south-east corner which is set to $-1$. 
Moreover, 
\begin{equation} \bm{R} = \textbf{diag}\{{\bm I}_{\ell_T + \ell_1 +1 },-1 \}. \label{MatrixR}
\end{equation}

\begin{figure}[tb]
	\centering
	\includegraphics[width=\linewidth]{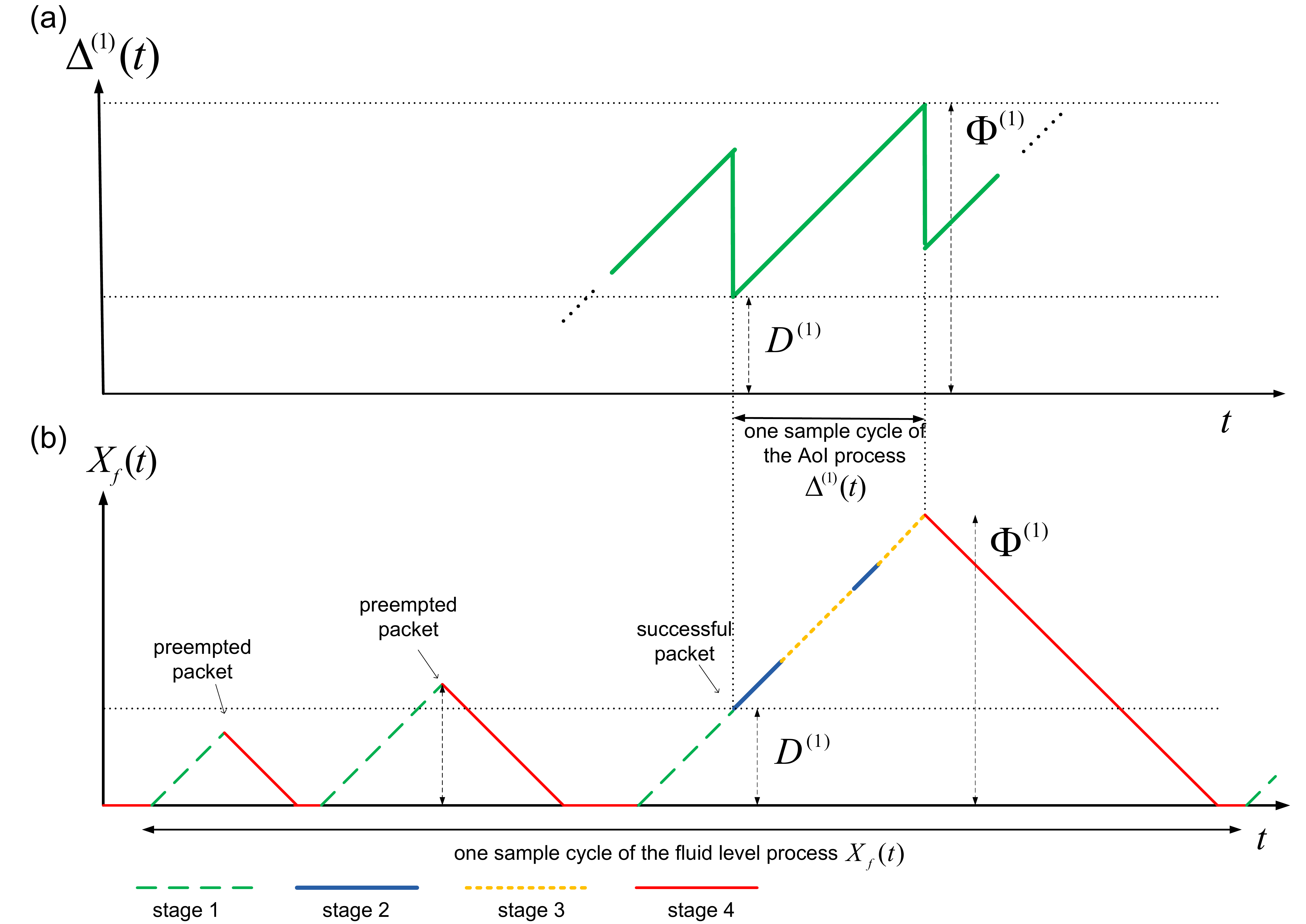}
	\caption{A sample path of the AoI process $\Delta^{(1)}(t)$ with emphasis on one of its sample cycles in given in subfigure (a) whereas subfigure (b) presents a sample path of the fluid level process $X_f(t)$ with emphasis on one sample cycle. Observe that the sample cycle of the former coincides with part of that of the latter.} 
	\label{fig:preemptive}
\end{figure}

Fig.~\ref{fig:preemptive}(a) illustrates one sample cycle of the AoI process $\Delta^{(1)}(t)$ which starts to rise from the value $D^{(1)}$ to the PAoI value $\Phi^{(1)}$. Fig.~\ref{fig:preemptive}(b) illustrates one sample cycle of the fluid level process $X_f(t)$ which takes the value $D^{(1)}$ at the epoch of transition from stage 1 to stage 2. The fluid level rises from the value $D^{(1)}$ to $\Phi^{(1)}$ during stages 2 and 3. Therefore, we observe that one sample cycle of the process $\Delta^{(1)}(t)$ coincides with part of the sample cycle of the process $X_f(t)$ associated with stages 2 and 3 only. Moreover, one sample value of the process $\Phi^{(1)}$ coincides with one sample value of the process $X_f(t)$ taken at the epoch of transition from stage 3 to stage 4.  
On the basis of these observations, we are now ready to state the following theorem which provides an expression for the pdfs of the AoI and PAoI processes in terms of the steady-state joint pdf vector $\bm{f(x)}$ of the MFQ $\bm{X(t)}$ given in the form \eqref{meform} with the matrix $\bm{A}$ being of size $b=\ell_T + \ell_1+1$ which is the number of positive drift states.
\begin{theorem}
	\label{theorem1}
	Consider the process ${\bm X(t)} \sim MFQ(\bm{Q},\bm{\tilde{Q}},\bm{R})$ with order $\ell_T + \ell_1 + 2$ with the characterizing matrices as defined in \eqref{MatrixQ} and \eqref{MatrixR} with 
	its steady-state joint pdf vector $\bm{f(x)}$ given in the form \eqref{meform}. Then, the pdf of the AoI process, $f_{\Delta^{(1)}}(x)$, and the pdf of the PAoI process,
	$f_{\Phi^{(1)}}(x)$, are given by the following closed form expression: 
	\begin{align}
	f_{\Delta^{(1)}}(x) & = \bm{g_A} e^{\bm{A}x} \bm{h_A} \ u(x),  \ \bm{h_A}= \bm{H} \left[ \begin{array}{c} \bm{0_{\ell_1 \times 1}} \\ 1 \\ \bm{1_{\ell_T}} \\ 0 \end{array} \right], \
	\bm{g_A} = \frac{1}{-\bm{g} \bm{A^{-1}} \bm{h_A}}, \label{AoIdist} \\
	f_{\Phi^{(1)}}(x) & = \bm{g_P} e^{\bm{A}x} \bm{h_P} \ u(x),  \ \bm{h_P}= \bm{H} \left[ \begin{array}{c} \bm{0_{(\ell_1 +1) \times 1}} \\ q_1 \bm{\nu^{(1)}} \\  
	\bm{0_{(\ell_T - \ell_1 +1) \times 1}}  \end{array} \right], \\ \nonumber 
	\bm{g_P} & = \frac{1}{-\bm{g} \bm{A^{-1}} \bm{h_P}},
	\label{PAoIdist}
	\end{align}
	Moreover, the associated non-central moments of the AoI and PAoI processes are given as follows: 
	\begin{align}
	E \left[(\Delta^{(1)})^i \right] &= (-1)^{i+1} \ i! \ \bm{g_A} \bm{A^{-(i+1)}} \bm{h_A}, i=1,2,\ldots,\\
	E \left[(\Phi^{(1)})^i \right] &= (-1)^{i+1} \ i! \ \bm{g_P} \bm{A^{-(i+1)}} \bm{h_P}, \  
	i=1,2,\ldots. \label{moments}
	\end{align}
\end{theorem}
The expression \eqref{AoIdist} stems from sample path arguments and requires censoring out the states in stages 1 and 4. This is achieved by the choice of $\bm{h_A}$ by which we sum up the joint pdfs in states belonging to stages 2 and 3 only. The choice of $\bm{g_A}$ is for normalization. 
The expression \eqref{AoIdist} similarly follows sample path arguments and requires the pdf of the fluid level just at the epoch of a transition from a state in stage 3 to state -1 in stage 4. 
For this purpose, $\bm{h_P}$ is chosen so as to sum up the joint pdfs belonging to states in stage 3 that have a transition to state -1, namely the states $(1,j)$ in stage 3, according to the transition rate vector $q_1 \bm{\nu^{(1)}}$. Again, the choice of $\bm{g_P}$ is for normalization. 

\begin{remark}
	While obtaining the steady-state joint pdf vector $\bm{f(x)}$ of the MFQ $\bm{X(t)}$ through the algorithm defined in three steps in \eqref{step1}-\eqref{meform}, we need an orthogonal matrix $\bm{U}$ satisfying \eqref{step1}. Let $\bm{v}$ be a column vector of ones (of system size) except for the last entry which is minus one. 
	Note that $\bm{v}$ is a right eigenvector of the matrix $\bm{Q} \bm{R}^{-1}$. Also, let $\bm{w}$ be a column vector of zeros except for the first entry which is one. 
	Let $\bm{u}=\bm{v} - || \bm{v} ||_2 \bm{w}$. 
	Then, the symmetric orthogonal matrix $\bm{U}$ defined by $\bm{U}={\bm I} - \frac{2 \bm{uu^T}}{\bm{u^T u}}$ gives rise to the factorization \eqref{step1} with the matrix $\bm{\Psi}$ reducing to a scalar which is actually zero. This process is called the Householder transformation in \cite{golub.vanloan.1996} and significantly reduces the complexity of \eqref{step1}.	
\end{remark}
\begin{remark}
	In the most general case, the MFQ $\bm{X(t)}$ has a system size of $\ell_T + \ell_1+2$. However, the system size is significantly reduced in some important sub-cases. 
	For example, consider the global preemption case, i.e., $\bm{P} = \bm{1_N} \bm{1_N^T}$ and when the service time requirements of the sources are homogeneous, i.e., $\Theta^{(n)} \sim \Theta \sim PH(\bm{\sigma},\bm{S})$ for $n=1,\ldots,N$ with order $\ell$ and homogeneous transmission errors and retransmission policies, i.e., $e_n = e, \ r_n = r, \ n=1,\ldots,N.$ From the perspective of AoI and PAoI of the tagged source-$1$, one can solve an auxiliary two-source system where the second source in the auxiliary system stands for the superposition of all the sources indexed from 2 to $N$ in the original system and $\bm{P} = \bm{1_2}\bm{1_2^T}$. The arising MFQ will now have a system size of $3 \ell + 2$ which does not depend on the number of users $N$.  Similar reductions are possible for non-premptive systems
	with homogeneous service time requirements.
\end{remark}

\section{Numerical Results}
\subsection{Validation with Simulations}

\begin{figure}[]
	\centering
	\includegraphics[width=0.8\linewidth]{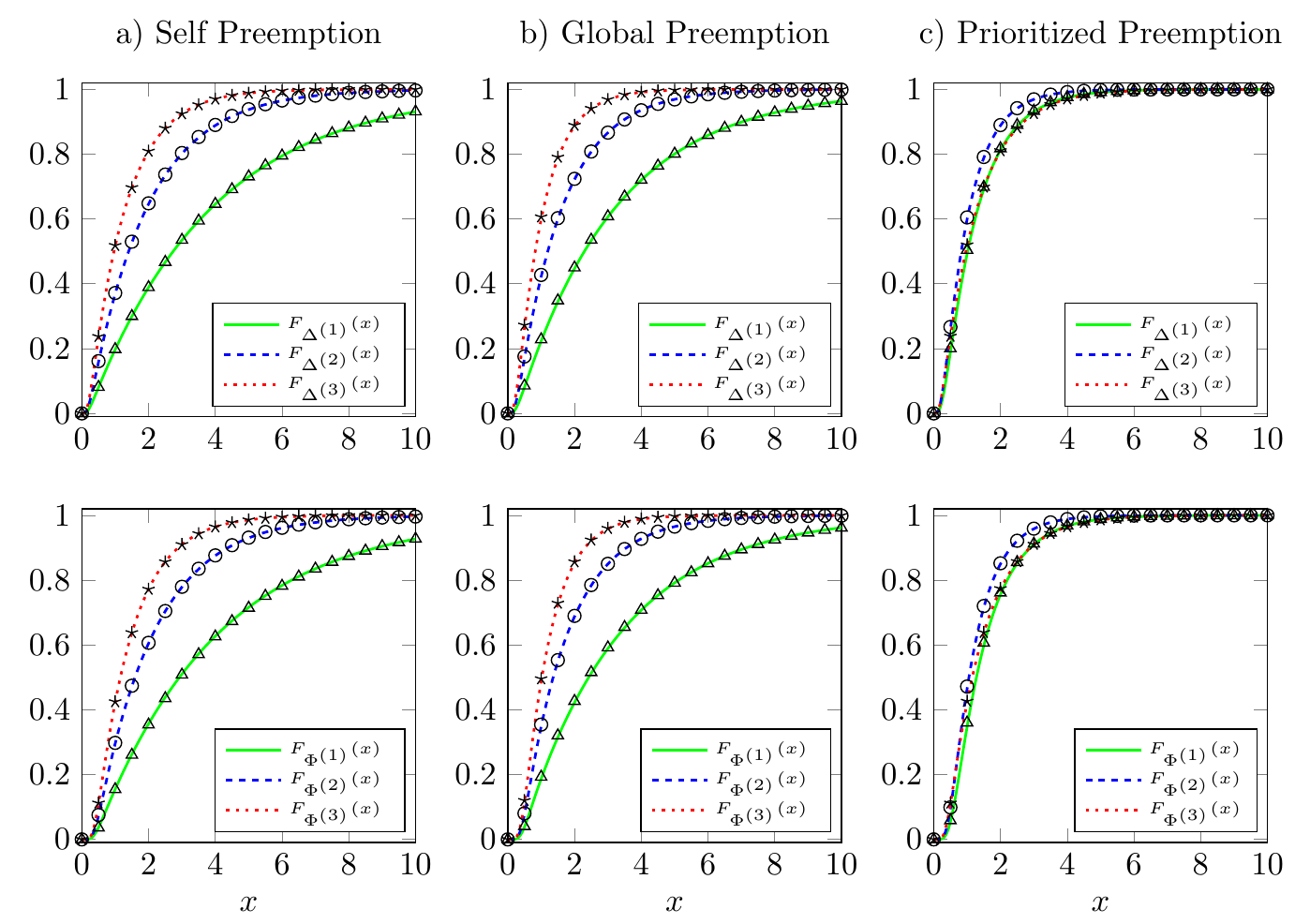}
	\caption{The cdf of both the AoI and PAoI processes of a system with 3 sources
		obtained by the proposed analytical method and simulations (shown by markers) for three cases (a) self preemption, (b) global preemption, (c) prioritized preemption, 
		when $(\lambda_1,\lambda_2,\lambda_3) = (1,2,3)$, $\rho=2/3$, and $c_{\Theta}^2 = 1/4$.}
	\label{fig:sim1}
\end{figure}
\begin{figure}[]
	\centering
	\includegraphics[width=0.8\linewidth]{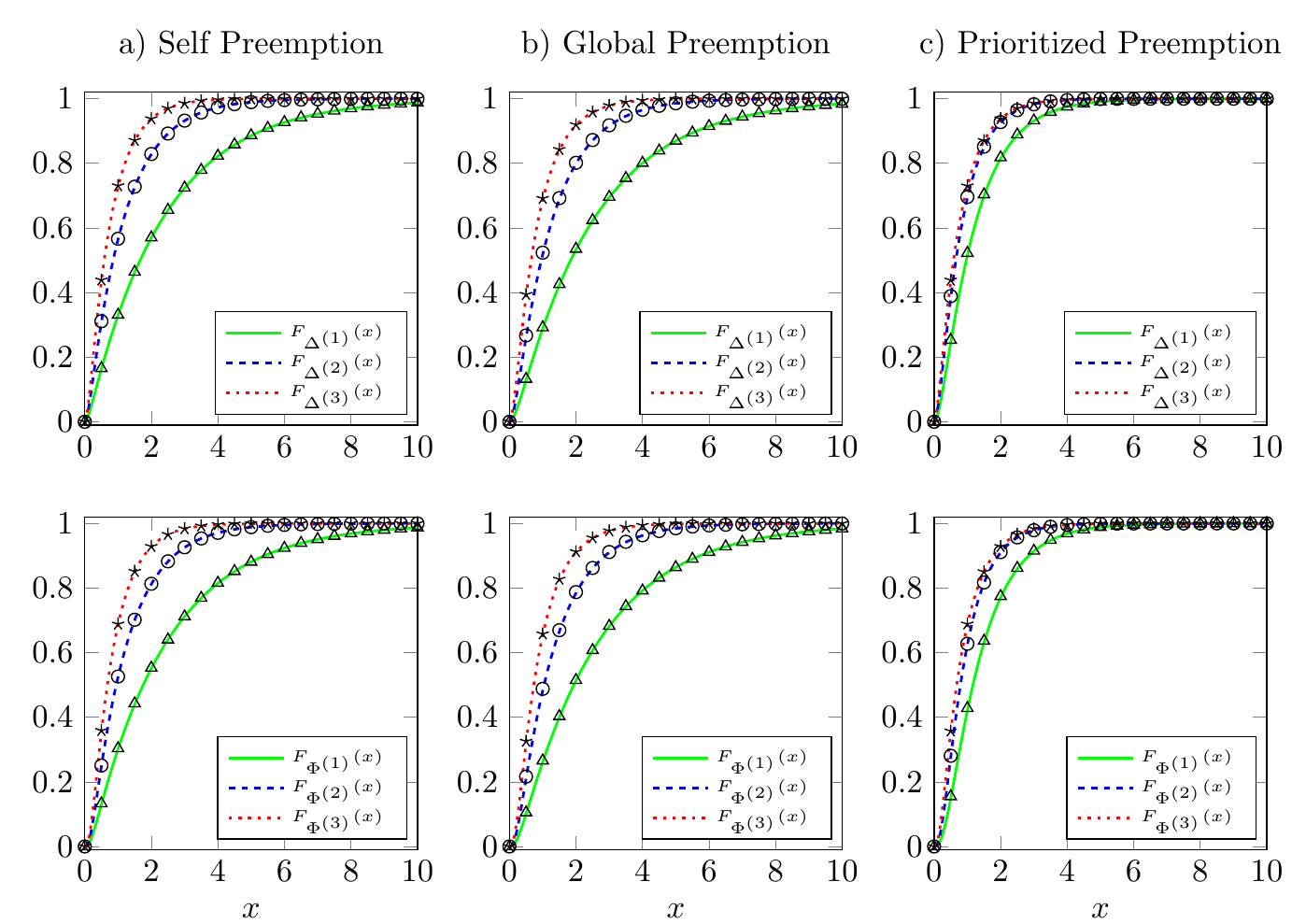}
	\caption{The cdf of both the AoI and PAoI processes of a system with 3 sources
		obtained by the proposed analytical method and simulations (shown by markers) for three cases (a) self preemption, (b) global preemption, (c) prioritized preemption, 
		when $(\lambda_1,\lambda_2,\lambda_3) = (1,2,3)$, $\rho=2/3$, and $c_{\Theta}^2 = 1/2$.
	}
	\label{fig:sim2}
\end{figure}
In the first set of numerical examples, we validate the proposed approach by comparing the obtained cdfs for AoI and PAoI processes against the empirical cdfs obtained with simulations in the context of a system with 3 sources with homogeneous service time requirements.
For this purpose, we fix the intensity vector $(\lambda_1,\lambda_2,\lambda_3)$ and for the service times, we use a PH-type distribution with mean $\rho/\lambda$ for given system load $\rho$ and the squared coefficient of variation of the service times is fixed to a given value $c_{\Theta}^2$ according the following procedure. 
For $c_{\Theta}^2=1/j \leq 1$ for a positive integer $j$, the $E(\mu^{-1},j)$ distribution is used which refers to an Erlang distribution with mean $\mu^{-1}$ and with order $j$. If $j$ is not an integer, then we resort to a mixture of two appropriate Erlang distributions \cite{tijms_book03}. When $c_{\Theta}^2 > 1$, then we propose to use a hyper-exponential distribution with balanced means to fit the first two moments \cite{tijms_book03}. In the first example, we fix $(\lambda_1,\lambda_2,\lambda_3)$ to $(1,2,3)$ and the system load $\rho$ to $2/3$. Moreover, we study three preemption policies: global preemption, self preemption, and prioritized preemption in which a newcoming class-$i$ packet preempts a class-$j$ packet in service if only if $i \leq j$, $\bm{P}$ is a lower-triangular matrix of ones at and below the main diagonal. The cdfs of the AoI and PAoI processes are depicted in figures~\ref{fig:sim1} and \ref{fig:sim2}, respectively, for the choice of $c_{\Theta}^2 = 1/4$ and $c_{\Theta}^2 = 1/2$, respectively. We have the following observations:
\begin{itemize}
	\item Perfect match with the simulation results are obtained in all cases.
	\item For symmetric preemption systems such as self preemption and global preemption, low intensity sources are penalized in terms of AoI and PAoI. However, with preferential treatment to low traffic intensity sources by means of non-symmetric policies such as prioritized preemption, this situation can be mitigated.
	\item When $c_{\Theta}^2$ increases, the overall performance of the system improves which can be inferred from the sharper increase of the cdf curves for all the three sources 
	in Fig.~\ref{fig:sim2} than in Fig.~\ref{fig:sim1}.
\end{itemize}
\subsection{Validation with Existing Results}
In this subsection, we will compare our findings with existing closed-form expressions in the existing literature for validation of the numerical accuracy of our proposed method.
As the first example, the reference \cite{yates_kaul_tit19} provides closed form expressions for the mean AoI for each of the sources in an $M/M/1/1$ system with homogeneous service time requirements, global preemption, and no transmission errors. For two different choices of the traffic intensity vector, and for various values of the load parameter $\rho$, the mean AoI values are tabulated in Table~\ref{table1} using the expressions in \cite{yates_kaul_tit19} and the proposed method for the case of global preemption. The two sets of results perfectly match up to four digits. For a similar $M/M/1/1$ system allowing transmission errors and employing the self preemption policy, the reference \cite{farazi_etal_Asilomar19} provides closed form expressions for the mean AoI for each of the sources. For two different choices of the traffic intensity vector, for various values of the load parameter $\rho$ and the transmission error parameter $e$ which is intact for each source, the mean AoI values are tabulated in Table~\ref{table2} using the expressions in \cite{farazi_etal_Asilomar19} and the proposed method for the case of self preemption with no retransmissions, i.e., $r_n=r=0, \ n=1,\ldots,N$. The two sets of results perfectly match up to four digits except for one single instance in which the match is up to three digits.
\begin{table*}[]
	\centering
	\caption{Mean AoI for each of the three sources obtained with the closed-form expressions in Ref.~\cite{yates_kaul_tit19} and the proposed method for 
		various choices of the traffic intensity vector and the load parameter $\rho$.}
	\begin{tabular}{ccccccccc}
		\hline 
		&  &  \multicolumn{2}{c}{$E[\Delta^{(1)}]$}& \multicolumn{2}{c}{$E[\Delta^{(2)}]$}& \multicolumn{2}{c}{$E[\Delta^{(3)}]$}  \\ \hline
		$(\lambda_1,\lambda_2,\lambda_3)$ & $\rho$    & Ref.~\cite{yates_kaul_tit19} & Proposed & Ref.~\cite{yates_kaul_tit19} & Proposed & Ref.~\cite{yates_kaul_tit19} & Proposed \\ \hline
		$(1,2,3)$ & 0.50 & 1.5000 & 1.5000 & 0.7500& 0.7500& 0.5000& 0.5000 \\ \cline{1-9}
		& 0.75  & 1.7500 & 1.7500 & 0.8750 & 0.8750& 0.5833& 0.5833 \\ \cline{2-9}
		& 1.00  & 2.0000 & 2.0000 & 1.0000 & 1.0000& 0.6667 & 0.6667 \\ \cline{2-9}
		& 1.25  & 2.2500& 2.2500& 1.1250& 1.1250& 0.7500& 0.7500\\ \cline{2-9}
		& 1.50  & 2.5000& 2.5000& 1.2500 & 1.2500& 0.8333 & 0.8333\\ \cline{2-9}
		$(1,4,16)$ & 0.50  & 1.5000 & 1.5000& 0.3750 &0.3750 & 0.0938 & 0.0938  \\ \cline{1-9}
		& 0.75  & 1.7500 & 1.7500 & 0.4375 & 0.4375& 0.1094 & 0.1094\\ \cline{2-9}
		& 1.00 & 2.0000 & 2.0000 & 0.5000& 0.5000& 0.1250 & 0.1250\\ \cline{2-9}
		& 1.25 & 2.2500& 2.2500& 0.5625& 0.5625& 0.1406 & 0.1406\\ \cline{2-9}
		& 1.50  & 2.5000& 2.5000& 0.6250 & 0.6250& 0.1563& 0.1563 \\ \hline
		
	\end{tabular}
	\label{table1}
\end{table*}

\begin{table*}[tb]
	\centering
	\caption{Mean AoI for each of the three sources obtained with the closed-form expressions in Ref.~\cite{farazi_etal_Asilomar19} and the proposed method for 
		various choices of the traffic intensity vector, the load parameter $\rho$, and the error parameter $e$ when the discarding parameter $r$ is set to zero.}
	\begin{tabular}{cccccccccccc}
		\hline 
		& & &   \multicolumn{2}{c}{$E[\Delta^{(1)}]$}&  \multicolumn{2}{c}{$E[\Delta^{(2)}]$}&  \multicolumn{2}{c}{$E[\Delta^{(3)}]$}  \\ \hline
		$(\lambda_1,\lambda_2,\lambda_3)$ & $\rho$ &$e$ & Ref.~\cite{farazi_etal_Asilomar19} & Proposed & Ref.~\cite{farazi_etal_Asilomar19} & Proposed & Ref.~\cite{farazi_etal_Asilomar19} & Proposed \\ \hline
		$(1,2,3)$ & 0.5 &0.04 & 1.5839 & 1.5839 & 0.7971 & 0.7971 & 0.5319 & 0.5319  \\ \cline{1-9}
		&     &0.10 & 1.6880 & 1.6880 & 0.8492 & 0.8492 & 0.5667 & 0.5667 \\ \cline{3-9}
		&     &0.25 & 2.0214 & 2.0214 & 1.0159 & 1.0159 & 0.6778 & 0.6778 \\ \cline{3-9}
		& 1.0 &0.04 & 2.1429 & 2.1429 & 1.0833 & 1.0833 & 0.7222 & 0.7222\\ \cline{2-9}
		&     &0.10 & 2.2817 & 2.2817 & 1.1528 & 1.1528 & 0.7685 & 0.7685\\ \cline{3-9}
		&     &0.25 & 2.7262 & 2.7262 & 1.3750 & 1.3750 & 0.9167 & 0.9167\\ \cline{3-9}
		& 1.5 &0.04 & 2.7042 & 2.7042 & 1.3688 & 1.3687 & 0.9109 & 0.9109 \\ \cline{2-9}
		&     &0.10 & 2.8778 & 2.8778 & 1.4556 & 1.4556 & 0.9688 & 0.9688 \\ \cline{3-9}
		&     &0.25 & 3.4333 & 3.4333 & 1.7333 & 1.7333 & 1.1540 & 1.1540\\ \cline{3-9}
		$(1,4,16)$ & 0.5 &0.04 & 1.5699 & 1.5699 & 0.3965 & 0.3965 & 0.0990 & 0.0990\\ \cline{1-9}
		&     &0.10 & 1.6740 & 1.6740 & 0.4225 & 0.4225 & 0.1055 & 0.1055\\ \cline{3-9}
		&     &0.25 & 2.0074 & 2.0074 & 0.5059 & 0.5059 & 0.1264 & 0.1264\\ \cline{3-9}
		& 1.0 &0.04 & 2.1050 & 2.1050 & 0.5370 & 0.5370 & 0.1334 & 0.1334\\ \cline{2-9}
		&     &0.10 & 2.2439 & 2.2439 & 0.5717 & 0.5717 & 0.1421 & 0.1421\\ \cline{3-9}
		&     &0.25 & 2.6883 & 2.6883 & 0.6829 & 0.6829 & 0.1699 & 0.1699\\ \cline{3-9}
		& 1.5 &0.04 & 2.6423 & 2.6423 & 0.6780 & 0.6780 & 0.1675 & 0.1675\\ \cline{2-9}
		&     &0.10 & 2.8159 & 2.8159 & 0.7214 & 0.7214 & 0.1784 & 0.1784 \\ \cline{3-9}
		&     &0.25 & 3.3714 & 3.3714 & 0.8603 & 0.8603 & 0.2131 & 0.2131\\ \hline

	\end{tabular}
	\label{table2}
\end{table*}
\subsection{Impact of Choice of Preemption Policies}
\begin{figure}[t]
	\centering
	\includegraphics[width=\linewidth]{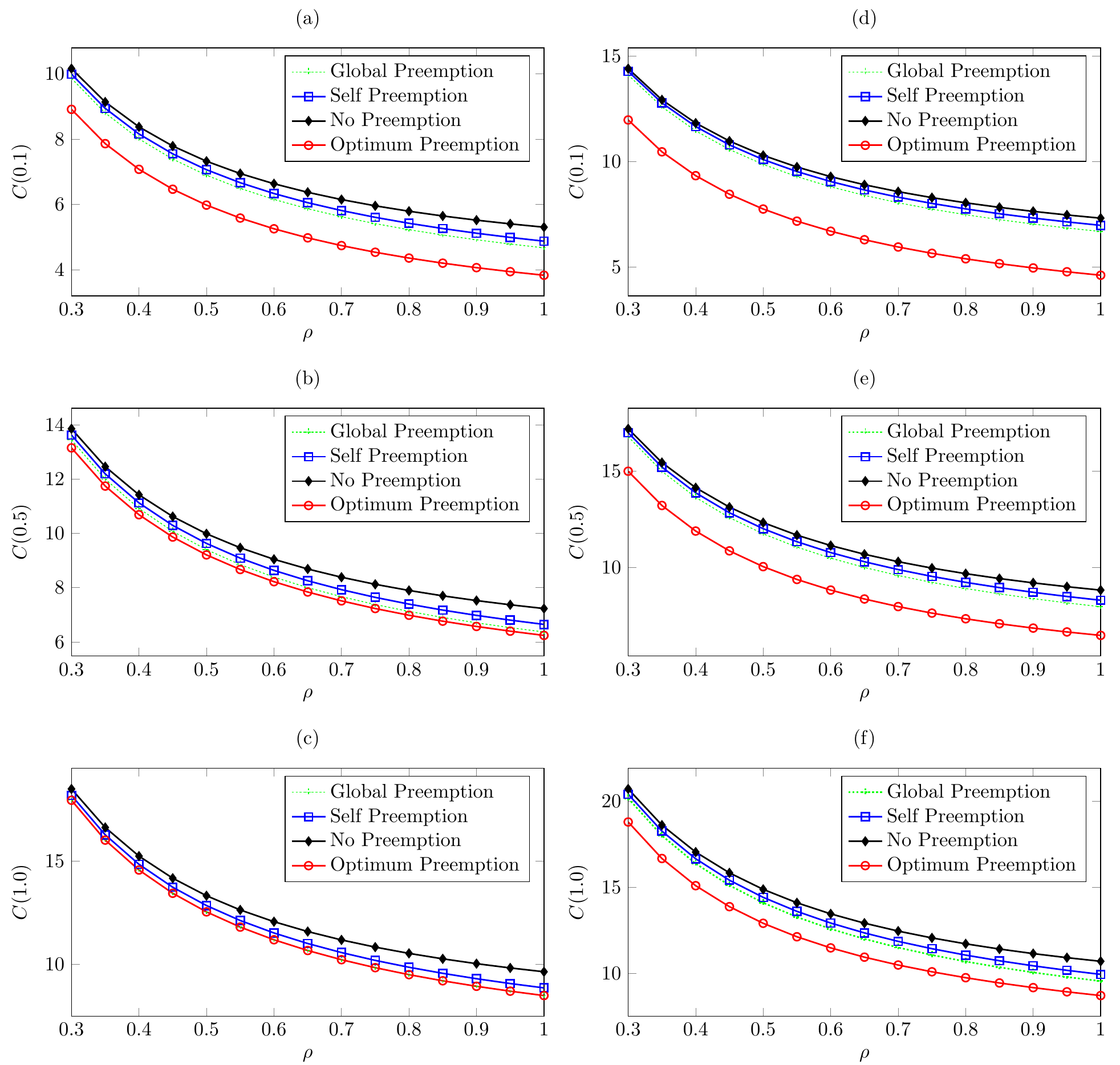}
	\caption{The cost function $C(\alpha)$ as a function of the load parameter $\rho$ for an $M/M/1/1$ system with four preemption policies when $\mu=1,r=0.9,e=0.1$ for three values of $\alpha \in \{ 0.1, 0.5, 1 \}$.  
		For the subfigures (a)-(c), $\rho_1=\rho_2$ whereas for the subfigures (d)-(f),  $\rho_2 = 2 \rho_1$.
	} 
	\label{fig:com1}
\end{figure}

In this subsection, we study the impact of the choice of preemption probabilities (using the analytical model only) on the system cost $C(\alpha)$ for a two-source preemptive $M/PH/1/1$ system with homogeneous service times, which is given in the following form:
\begin{equation}
C(\alpha)= E[\Delta^{(1)}] + \alpha E[\Delta^{(2)}], \ 0 \leq \alpha \leq 1, \label{cost}
\end{equation} 
which allows one to give more importance to source-$1$ over source-$2$ with a proper choice of the cost parameter $\alpha$. When $\alpha=1$, both sources are equally important whereas when $\alpha=0$, the age of the second source is irrelevant. 
For a given traffic intensity vector, load $\rho$, and $c_{\Theta}^2$ for the homogeneous service times, we employ a preemption matrix $\bm P$ such that $P_{1,1} = P_{2,2} = P_d$ and we perform brute-force optimization to find the optimum choices of $P_d$, $P_{1,2}$, and $P_{2,1}$, denoted by $P_d^{\ast}$, $P_{1,2}^{\ast}$, and $P_{2,1}^{\ast}$, respectively, so that the cost function in \eqref{cost} is minimized for a given cost parameter $\alpha$ (with a resolution of $0.05$ for each parameter). Table~\ref{table3} provides our findings. We have the following observations:
\begin{itemize}
	\item The optimum value $P_d^{\ast}$ appears to depend on $c_{\Theta}^2$ and not on the specific choices of $\alpha$ and the traffic mix. When $c_{\Theta}^2$ is small (large), $P_d^{\ast}$ is zero (one).
	\item The optimum values $P_{1,2}^{\ast}$ and $P_{2,1}^{\ast}$ appear to depend on $c_{\Theta}^2$, $\alpha$, and the traffic mix.  All other input parameters being fixed,  $P_{2,1}^{\ast}$ decreases and $P_{1,2}^{\ast}$ increases with increased cost parameter $\alpha$ but their actual values themselves depend on the other parameters.  
\end{itemize}
\begin{table*}[thb]
	\centering
	\caption{Optimum preemption parameters $P_d^{\ast}$, $P_{1,2}^{\ast}$, and $P_{2,1}^{\ast}$, which minimize $C(\alpha)$ 
		for various values of traffic intensity vector, $c_{\Theta}^2$, and $\alpha$.}
	\begin{tabular}{cc|ccc|ccc|ccc}
		\hline 
		& &  \multicolumn{9}{c}{($\lambda_1,\lambda_2$)} \\
		\hline
		$c_\Theta^{2}$  &$\alpha$  &  \multicolumn{3}{c|}{$(1,2)$} & \multicolumn{3}{c|}{$(1,1)$} & \multicolumn{3}{c}{$(2,1)$}\\ \cline{3-11}
		&    & $P_{d}^{\ast}$ & $P_{2,1}^{\ast}$ & $P_{1,2}^{\ast}$& $P_{d}^{\ast}$ & $P_{2,1}^{\ast}$ & $P_{1,2}^{\ast}$& $P_{d}^{\ast}$ & $P_{2,1}^{\ast}$ & $P_{1,2}^{\ast}$ \\ \hline 
		1/16 & 0.25 & 0.00  & 1.00 & 0.00& 0.00& 0.65& 0.00& 0.00& 0.00& 0.00\\ \cline{2-11}
		& 0.50 & 0.00  & 1.00 & 0.00& 0.00& 0.30& 0.00& 0.00& 0.00& 0.25\\ \cline{2-11}
		& 0.75 & 0.00  & 0.90 & 0.00& 0.00& 0.05& 0.00& 0.00& 0.00& 0.50\\ \cline{2-11}
		& 1.00 & 0.00  & 0.70 & 0.00& 0.00& 0.00& 0.00& 0.00& 0.00& 0.70\\ \cline{1-11}
		1/4  & 0.25 & 0.00  & 1.00 & 0.00& 0.00& 0.75& 0.00& 0.00& 0.00& 0.00\\ \cline{2-11}
		& 0.50 & 0.00  & 1.00 & 0.00& 0.00& 0.30& 0.00& 0.00& 0.00& 0.30\\ \cline{2-11}
		& 0.75 & 0.00  & 0.95 & 0.00& 0.00& 0.10& 0.00& 0.00& 0.00& 0.55\\ \cline{2-11}
		& 1.00 & 0.00  & 0.75 & 0.00& 0.00& 0.00& 0.00& 0.00& 0.00& 0.75\\ \cline{1-11}
		1  & 0.25 & 1.00  & 1.00 & 0.00& 1.00& 1.00& 0.15& 1.00& 1.00& 1.00\\ \cline{2-11}
		& 0.50 & 1.00  & 1.00 & 0.00& 1.00& 1.00& 0.55& 1.00& 0.60& 1.00\\ \cline{2-11}
		& 0.75 & 1.00  & 1.00 & 0.05& 1.00& 1.00& 0.85& 1.00& 0.35& 1.00\\ \cline{2-11}
		& 1.00 & 1.00  & 1.00 & 0.15& 1.00& 1.00& 1.00& 1.00& 0.15& 1.00\\ \cline{1-11}

	\end{tabular}
	\label{table3}
\end{table*}
Finally, we depict the cost $C(\alpha)$ as a function of the system load $\rho$ for four preemption 
policies, namely non-preemptive, self-preemptive, globally preemptive, and optimum 
preemptive policies (as obtained using the brute-force approach outlined above) in 
Fig.~\ref{fig:com1}. We observe that optimum preemption significantly outperforms
all the other policies with the level of performance improvement increases with decreased
cost parameter $\alpha$. When $\alpha=1$ and the traffic mix is even, global preemption
and optimum preemption yield the same performance.  When $\alpha=1$ and the traffic mix 
is not even, i.e., $\rho_2 = 2 \rho_1$, then there is still a substantial gain attained 
with the optimum preemptive policy when compared with the other preemption policies. 
For homogeneous exponential service times, we observe that global preemption performs 
better than self preemption which also outperforms the non-preemptive policy for all 
the cases we studied.

\section{Conclusions}
In this paper, we propose a novel method to obtain the exact distributions of the AoI and PAoI for a  probabilistically preemptive bufferless heterogeneous $M/PH/1/1$ queueing system with packet errors using Markov fluid queues. This model is more general than many recent existing models arising in status update systems and obtaining the distributions in addition to mean values is the major contribution of this paper.
Numerical examples are provided to validate the proposed approach and its numerical accuracy. We also provide a number of examples for which probabilistic preemption is substantially beneficial when compared to conventional non-preemptive, self-preemptive, or globally-preemptive policies in terms of AoI. Such probabilistic schemes can also be used to provide source differentiation in multi-source status update systems.

\bibliographystyle{ieeetran}
\bibliography{AoIBufferless}

\begin{thebibliography}{10}
\providecommand{\url}[1]{#1}
\csname url@samestyle\endcsname
\providecommand{\newblock}{\relax}
\providecommand{\bibinfo}[2]{#2}
\providecommand{\BIBentrySTDinterwordspacing}{\spaceskip=0pt\relax}
\providecommand{\BIBentryALTinterwordstretchfactor}{4}
\providecommand{\BIBentryALTinterwordspacing}{\spaceskip=\fontdimen2\font plus
\BIBentryALTinterwordstretchfactor\fontdimen3\font minus
  \fontdimen4\font\relax}
\providecommand{\BIBforeignlanguage}[2]{{%
\expandafter\ifx\csname l@#1\endcsname\relax
\typeout{** WARNING: IEEEtran.bst: No hyphenation pattern has been}%
\typeout{** loaded for the language `#1'. Using the pattern for}%
\typeout{** the default language instead.}%
\else
\language=\csname l@#1\endcsname
\fi
#2}}
\providecommand{\BIBdecl}{\relax}
\BIBdecl

\bibitem{kaul_etal_SMAN11}
S.~{Kaul}, M.~{Gruteser}, V.~{Rai}, and J.~{Kenney}, ``Minimizing age of
  information in vehicular networks,'' in \emph{2011 8th Annual IEEE
  Communications Society Conference on Sensor, Mesh and Ad Hoc Communications
  and Networks}, June 2011, pp. 350--358.

\bibitem{kaul_etal_infocom12}
S.~{Kaul}, R.~{Yates}, and M.~{Gruteser}, ``Real-time status: How often should
  one update?'' in \emph{2012 Proceedings IEEE INFOCOM}, March 2012, pp.
  2731--2735.

\bibitem{kaul_etal_ciss12}
S.~K. {Kaul}, R.~D. {Yates}, and M.~{Gruteser}, ``Status updates through
  queues,'' in \emph{2012 46th Annual Conference on Information Sciences and
  Systems (CISS)}, March 2012, pp. 1--6.

\bibitem{pappas_etal_icc15}
N.~{Pappas}, J.~{Gunnarsson}, L.~{Kratz}, M.~{Kountouris}, and V.~{Angelakis},
  ``Age of information of multiple sources with queue management,'' in
  \emph{2015 IEEE International Conference on Communications (ICC)}, June 2015,
  pp. 5935--5940.

\bibitem{kosta_etal}
A.~{Kosta}, N.~{Pappas}, A.~{Ephremides}, and V.~{Angelakis}, ``Age and value
  of information: Non-linear age case,'' in \emph{2017 IEEE International
  Symposium on Information Theory (ISIT)}, June 2017, pp. 326--330.

\bibitem{kosta_etal_survey}
A.~Kosta, N.~Pappas, and V.~Angelakis, ``Age of information: A new concept,
  metric, and tool,'' \emph{Foundations and Trends® in Networking}, vol.~12,
  no.~3, pp. 162--259, 2017.

\bibitem{sun_etal_tit17}
Y.~{Sun}, E.~{Uysal-Biyikoglu}, R.~D. {Yates}, C.~E. {Koksal}, and N.~B.
  {Shroff}, ``Update or wait: How to keep your data fresh,'' \emph{IEEE
  Transactions on Information Theory}, vol.~63, no.~11, pp. 7492--7508, Nov
  2017.

\bibitem{yates_kaul_tit19}
R.~D. {Yates} and S.~K. {Kaul}, ``The age of information: Real-time status
  updating by multiple sources,'' \emph{IEEE Transactions on Information
  Theory}, vol.~65, no.~3, pp. 1807--1827, 2019.

\bibitem{inoue_etal_tit19}
Y.~{Inoue}, H.~{Masuyama}, T.~{Takine}, and T.~{Tanaka}, ``A general formula
  for the stationary distribution of the age of information and its application
  to single-server queues,'' \emph{IEEE Transactions on Information Theory},
  pp. 1--1, 2019.

\bibitem{costa_etal_TIT16}
M.~{Costa}, M.~{Codreanu}, and A.~{Ephremides}, ``On the age of information in
  status update systems with packet management,'' \emph{IEEE Transactions on
  Information Theory}, vol.~62, no.~4, pp. 1897--1910, April 2016.

\bibitem{chen_huang_isit16}
K.~{Chen} and L.~{Huang}, ``Age-of-information in the presence of error,'' in
  \emph{2016 IEEE International Symposium on Information Theory (ISIT)}, July
  2016, pp. 2579--2583.

\bibitem{huang_modiano}
L.~{Huang} and E.~{Modiano}, ``Optimizing age-of-information in a multi-class
  queueing system,'' in \emph{2015 IEEE International Symposium on Information
  Theory (ISIT)}, June 2015, pp. 1681--1685.

\bibitem{arafa_ulukus_asilomar17}
A.~{Arafa} and S.~{Ulukus}, ``Age minimization in energy harvesting
  communications: Energy-controlled delays,'' in \emph{2017 51st Asilomar
  Conference on Signals, Systems, and Computers}, Oct 2017, pp. 1801--1805.

\bibitem{hsu_etal_isit17}
Y.~{Hsu}, E.~{Modiano}, and L.~{Duan}, ``Age of information: Design and
  analysis of optimal scheduling algorithms,'' in \emph{2017 IEEE International
  Symposium on Information Theory (ISIT)}, June 2017, pp. 561--565.

\bibitem{he_etal_TIT18}
Q.~{He}, D.~{Yuan}, and A.~{Ephremides}, ``Optimal link scheduling for age
  minimization in wireless systems,'' \emph{IEEE Transactions on Information
  Theory}, vol.~64, no.~7, pp. 5381--5394, July 2018.

\bibitem{costa_peak}
M.~{Costa}, M.~{Codreanu}, and A.~{Ephremides}, ``Age of information with
  packet management,'' in \emph{2014 IEEE International Symposium on
  Information Theory}, 2014, pp. 1583--1587.

\bibitem{neuts81}
M.~F. Neuts, \emph{Matrix-geometric Solutions in Stochastic Models: An
  Algorithmic Approach}.\hskip 1em plus 0.5em minus 0.4em\relax Dover
  Publications, Inc., 1981.

\bibitem{anick_mitra82}
D.~Anick, D.~Mitra, and M.~M. Sondhi, ``Stochastic theory of a data-handling
  system with multiple sources,'' \emph{Bell System Technical Journal},
  vol.~61, no.~8, pp. 1871--1894, 1982.

\bibitem{kosten.1984}
L.~Kosten, ``Stochastic theory of data handling systems with groups of multiple
  sources,'' \emph{Performance of Computer Communication Systems}, pp.
  321--331, 1984.

\bibitem{kulkarni_1997}
V.~G. Kulkarni, ``Fluid models for single buffer systems,'' in \emph{Frontiers
  in Queueing: Models and Applications in Science and Engineering}, J.~H.
  Dshalalow, Ed.\hskip 1em plus 0.5em minus 0.4em\relax Boca Raton, FL, USA:
  CRC Press, Inc., 1997, ch. Fluid Models for Single Buffer Systems, pp.
  321--338.

\bibitem{yates_kaul_ISIT12}
R.~D. {Yates} and S.~{Kaul}, ``Real-time status updating: Multiple sources,''
  in \emph{2012 IEEE International Symposium on Information Theory
  Proceedings}, July 2012, pp. 2666--2670.

\bibitem{inoue_etal_ISIT17}
Y.~{Inoue}, H.~{Masuyama}, T.~{Takine}, and T.~{Tanaka}, ``The stationary
  distribution of the age of information in {FCFS} single-server queues,'' in
  \emph{2017 IEEE International Symposium on Information Theory (ISIT)}, June
  2017, pp. 571--575.

\bibitem{najm_nasser_isit16}
E.~{Najm} and R.~{Nasser}, ``Age of information: The gamma awakening,'' in
  \emph{2016 IEEE International Symposium on Information Theory (ISIT)}, July
  2016, pp. 2574--2578.

\bibitem{akar_etal_tcom20}
N.~{Akar}, O.~{Dogan}, and E.~U. {Atay}, ``Finding the exact distribution of
  (peak) age of information for queues of {PH/PH/1/1 and M/PH/1/2} type,''
  2020, to appear at IEEE Trans. Commun.

\bibitem{kosta_etal_isit19}
A.~{Kosta}, N.~{Pappas}, A.~{Ephremides}, and V.~{Angelakis}, ``Queue
  management for age sensitive status updates,'' in \emph{2019 IEEE
  International Symposium on Information Theory (ISIT)}, 2019, pp. 330--334.

\bibitem{soysal_ulukus_unpublished}
A.~Soysal and S.~Ulukus, ``Age of information in {G/G/1/1} systems: Age
  expressions, bounds, special cases, and optimization,'' \emph{CoRR}, vol.
  abs/1905.13743, 2019.

\bibitem{champati_etal_infocom19}
J.~P. {Champati}, H.~{Al-Zubaidy}, and J.~{Gross}, ``On the distribution of
  {AoI} for the {GI/GI/1/1} and {GI/GI/1/2*} systems: {Exact} expressions and
  bounds,'' in \emph{IEEE INFOCOM 2019 - IEEE Conference on Computer
  Communications}, April 2019, pp. 37--45.

\bibitem{najm2018status}
E.~Najm and E.~Telatar, ``Status updates in a multi-stream m/g/1/1 preemptive
  queue,'' 2018.

\bibitem{farazi_etal_Asilomar19}
S.~{Farazi}, A.~G. {Klein}, and D.~{Richard Brown}, ``Average age of
  information in multi-source self-preemptive status update systems with packet
  delivery errors,'' in \emph{2019 53rd Asilomar Conference on Signals,
  Systems, and Computers}, 2019, pp. 396--400.

\bibitem{moltafet2020average}
M.~Moltafet, M.~Leinonen, and M.~Codreanu, ``Average age of information for a
  multi-source {M/M/1} queueing model with packet management,'' 2020.

\bibitem{kaul2020timely}
S.~K. Kaul and R.~D. Yates, ``Timely updates by multiple sources: {The M/M/1}
  queue revisited,'' 2020.

\bibitem{yates_etal_isit19}
R.~D. {Yates}, J.~{Zhong}, and W.~{Zhang}, ``Updates with multiple service
  classes,'' in \emph{2019 IEEE International Symposium on Information Theory
  (ISIT)}, 2019, pp. 1017--1021.

\bibitem{asmussen_etal_SJS96}
S.~Asmussen, O.~Nerman, and M.~Olsson, ``Fitting phase-type distributions via
  the em algorithm,'' \emph{Scandinavian Journal of Statistics}, vol.~23,
  no.~4, pp. 419--441, 1996.

\bibitem{ocinneide}
C.~A. O’Cinneide, ``Characterization of phase-type distributions,''
  \emph{Communications in Statistics. Stochastic Models}, vol.~6, no.~1, pp.
  1--57, 1990.

\bibitem{PhFit}
A.~Horv{\'a}th and M.~Telek, ``{PhFit:} a general phase-type fitting tool,'' in
  \emph{Computer Performance Evaluation: Modelling Techniques and Tools},
  T.~Field, P.~G. Harrison, J.~Bradley, and U.~Harder, Eds.\hskip 1em plus
  0.5em minus 0.4em\relax Berlin, Heidelberg: Springer Berlin Heidelberg, 2002,
  pp. 82--91.

\bibitem{HiroyukiOkamura2016}
H.~Okamura and T.~Dohi, ``Ph fitting algorithm and its application to
  reliability engineering,'' \emph{Journal of the Operations Research Society
  of Japan}, vol.~59, no.~1, pp. 72--109, 2016.

\bibitem{akar_sohraby_jap04}
N.~Akar and K.~Sohraby, ``Infinite- and finite-buffer {Markov} fluid queues: a
  unified analysis,'' \emph{J. Appl. Probab.}, vol.~41, no.~2, pp. 557--569, 06
  2004.

\bibitem{golub.vanloan.1996}
G.~H. Golub and C.~F. van Loan, \emph{Matrix Computations}.\hskip 1em plus
  0.5em minus 0.4em\relax The Johns Hopkins University Press, 1996.

\bibitem{kankaya.2008}
H.~E. Kankaya and N.~Akar, ``{Solving multi-regime feedback fluid queues},''
  \emph{{Stochastic Models}}, vol.~24, no.~3, pp. 425--450, 2008.

\bibitem{soares_latouche}
A.~da~Silva~Soares and G.~Latouche, ``Fluid queues with level dependent
  evolution,'' \emph{European Journal of Operational Research}, vol. 196,
  no.~3, pp. 1041 -- 1048, 2009.

\bibitem{horvath_vanhoudt}
G.~{Horvath} and B.~{Van Houdt}, ``A multi-layer fluid queue with boundary
  phase transitions and its application to the analysis of multi-type queues
  with general customer impatience,'' in \emph{2012 Ninth International
  Conference on Quantitative Evaluation of Systems}, 2012, pp. 23--32.

\bibitem{tijms_book03}
H.~C. Tijms, \emph{A First Course in Stochastic Models}.\hskip 1em plus 0.5em
  minus 0.4em\relax West Sussex, England: John Wiley \& Sons, Inc., 2003.

\end{thebibliography}
\end{document}